\documentclass[useAMS,usenatbib]{mn2e}

\usepackage{graphicx}

 \def\hii{H\,\textsc{ii}}

\title[Cepheids in Leo~A]{The ACS LCID Project. VIII. The short-period
    Cepheids of Leo~A\thanks{Based on observations made with the NASA/ESA
    {\it Hubble Space Telescope}, obtained at the Space Telescope Science
    Institute, which is operated by the Association of Universities for
    Research in Astronomy, Inc., under NASA contract NAS5-26555. These
    observations are associated with program 10590.}}
\author[E.~J.\ Bernard et al.]
{Edouard J.\ Bernard,$^{1}$\thanks{E-mail: ejb@roe.ac.uk}
Matteo Monelli,$^{2,3}$
Carme Gallart,$^{2,3}$
Giuliana Fiorentino,$^{4}$
\newauthor
Santi Cassisi,$^{5}$
Antonio Aparicio,$^{2,3}$
Andrew A. Cole,$^{6}$
Igor Drozdovsky,$^{2,3,7}$
\newauthor
Sebastian L.\ Hidalgo,$^{2,3}$
Evan D.\ Skillman,$^{8}$
Peter B.\ Stetson,$^{9}$
Eline Tolstoy$^{10}$ \\
$^{1}$SUPA, Institute for Astronomy, University of Edinburgh, Royal
      Observatory, Blackford Hill, Edinburgh EH9 3HJ \\
$^{2}$Instituto de Astrof\'{i}sica de Canarias, Calle V\'ia L\'actea s/n, 38205 La Laguna, Tenerife, Spain \\
$^{3}$Departamento de Astrof\'{i}sica, Universidad de La Laguna, 38200 Tenerife, Spain \\
$^{4}$INAF-Osservatorio Astronomico di Bologna, via Ranzani 1, 40127 Bologna, Italy \\
$^{5}$INAF-Osservatorio Astronomico di Teramo, via M. Maggini, 64100 Teramo, Italy \\
$^{6}$School of Mathematics \& Physics, University of Tasmania, Hobart, Tasmania, Australia \\
$^{7}$Astronomical Institute, St. Petersburg State University, St. Petersburg, Russia \\
$^{8}$Minnesota Institute for Astrophysics, University of Minnesota, Minneapolis, MN 55455, USA \\
$^{9}$NRC Herzberg Institute for Astrophysics, 5071 West Saanich Road, Victoria, BC V9E 2E7, Canada \\
$^{10}$Kapteyn Astronomical Institute, University of Groningen,
    Groningen, Netherlands}

\begin{document}

\date{Accepted 2013 April 16. Received 2013 April 1; in original form 2012 December 29}

\pagerange{\pageref{firstpage}--\pageref{lastpage}} \pubyear{2012}

\maketitle

\label{firstpage}

\begin{abstract}

 We present the results of a new search for variable stars in the Local Group
 dwarf galaxy Leo~A, based on deep photometry from the Advanced
 Camera for Surveys onboard the {\it Hubble Space Telescope}. We detected 166
 bona fide variables in our field, of which about 60 percent are new discoveries,
 and 33 candidate variables. Of the confirmed variables, we found 156 Cepheids,
 but only 10 RR~Lyrae stars despite nearly 100 percent completeness at the
 magnitude of the horizontal branch. The RR~Lyrae stars include 7 fundamental
 and 3 first-overtone pulsators, with mean periods of 0.636 and 0.366 day,
 respectively.
 From their position on the period-luminosity (PL) diagram and light-curve
 morphology, we classify 91, 58, and 4 Cepheids as fundamental, first-overtone,
 and second-overtone mode Classical Cepheids (CC), respectively, and two as
 population II Cepheids. However, due to the low metallicity of Leo~A, about 90
 percent of the detected Cepheids have periods shorter than 1.5~days.
 Comparison with theoretical models indicate that some of the fainter stars
 classified as CC could be Anomalous Cepheids. We estimate the distance to Leo~A
 using the tip of the RGB (TRGB) and various methods based on the photometric
 and pulsational properties of the Cepheids and RR~Lyrae stars. The distances
 obtained with the TRGB and RR~Lyrae stars agree well with each other while that
 from the Cepheid PL relations is somewhat larger, which may indicate a mild
 metallicity effect on the luminosity of the short-period Cepheids. Due to its
 very low metallicity, Leo~A thus serves as a valuable calibrator of the
 metallicity dependencies of the variable star luminosities.

\end{abstract}


   \defcitealias{ber09}{Paper~I}
   \defcitealias{ber10}{Paper~II}
   \defcitealias{dol02}{D02}
   \defcitealias{goe07}{G07}
   \defcitealias{hoe94}{H94}
   \defcitealias{sch02}{Schulte-Ladbeck et al.\ 2002}


\section{Introduction}

 Classical Cepheids (CC) are the prime distance estimators in the nearby
 universe thanks to the \citet{lea08} law, which relates their brightness and
 period. However, their use is complicated by the existence of other types of
 Cepheids with different luminosities at a given period: population II Cepheids
 (P2C) and Anomalous Cepheids (AC). While the former have periods similar to
 those of CC (from less than a day to more than 100 days), they are typically
 fainter at a given period ($\sim$1$-$2~mag) such that recognizing CC and P2C
 is usually straightforward. On the other hand, AC are only slightly fainter
 than short-period CC \citep[see, e.g.][for a discussion of their formation and
 evolution]{bon97}. This confusion is even more significant for metal-poor
 galaxies, in which most of the CC have periods similar to AC
 \citep[$\sim$0.4$-$3~days; e.g.\ IC\,1613, Sextans A,
 Phoenix:][]{dol01,dol03,gal04,ber10}.

\begin{figure}
 \includegraphics[width=8.3cm]{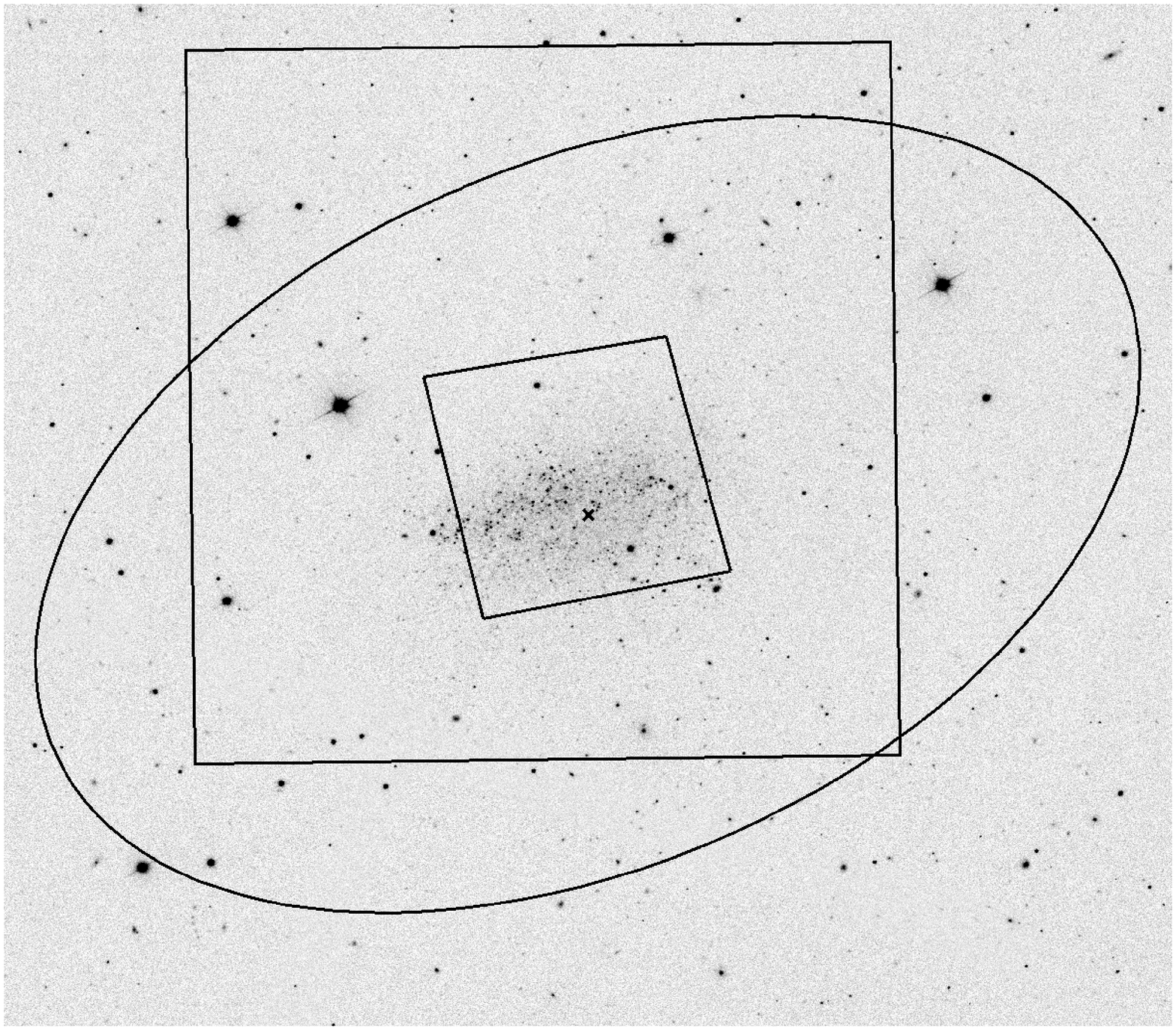}
 \caption{Sloan Digital Sky Survey image of Leo~A. The cross and ellipse
 indicate the center and extent of the galaxy \citep[R.A.=09$^h$59$^m$24.0$^s$,
 decl.=30$\degr$44$\arcmin$47$\arcsec$, $r=8\arcmin$;][]{van04}, while the small
 and large squares show the location of our field-of-view and that of
 \citet{dol02}, respectively.}
 \label{fig:loc}
\end{figure}

 An increasing number of studies also suggest that there may be a metallicity
 effect on the slope and/or zero-point of the period-luminosity (PL) relations
 \citep[e.g.][]{sak04,rom08,san09,bon10,sto11},
 which would affect the distance measurements of galaxies for which metallicity
 is not well constrained. It is therefore essential to obtain complete surveys
 of Cepheids in a sample of metal-poor galaxies in order to gain a
 better understanding of their physical and evolutionary properties.

 Leo~A, a small star forming dwarf galaxy discovered by \citet{zwi42} and
 located about 790~kpc away \citep[e.g.][hereinafter D02]{dol02}, is among the
 most metal-poor galaxies of the Local Group (LG). Its current metallicity is
 A$_{\rm O}$=12+log(O/H)=7.30$\pm$0.05 (i.e.\ [Fe/H]~$\sim-$1.5, or $\sim$3\%
 solar assuming A$_{\rm O,\sun}$=8.87$\pm$0.07; \citealt{gre93}) from the
 spectroscopy of three \hii\ regions and a planetary nebula \citep{ski89,van06}.
 \citet{tol98} drew attention to the relative lack of an ancient stellar
 population in Leo~A, and this was supported by the discovery of the rarity of
 RR Lyrae stars by \citetalias{dol02}.

\begin{table}
 \centering
 \begin{minipage}{130mm}
  \caption{Summary of Observations.\label{tab:tab1}}
  \begin{tabular}{@{}ccccc@{}}
  \hline
Date & UT Start & MHJD\footnote{Modified Heliocentric Julian Date of mid-exposure: HJD$-$2,400,000.} & Filter & Exp. Time (s) \\
  \hline
 2005-12-26 & 08:45:32 & 53730.369406 & F475W & 1200  \\
 2005-12-26 & 09:07:48 & 53730.384869 & F475W & 1200  \\
 2005-12-26 & 10:18:55 & 53730.434371 & F814W & 1220  \\
 2005-12-26 & 10:41:31 & 53730.450066 & F814W & 1220  \\
 2005-12-26 & 19:56:55 & 53730.835645 & F475W & 1200  \\
 2005-12-26 & 20:19:11 & 53730.851108 & F475W & 1200  \\
 2005-12-26 & 21:30:18 & 53730.900610 & F814W & 1220  \\
 2005-12-26 & 21:52:54 & 53730.916305 & F814W & 1220  \\
 2005-12-28 & 19:54:20 & 53732.833851 & F475W & 1200  \\
 2005-12-28 & 20:16:36 & 53732.849314 & F475W & 1200  \\
 2005-12-28 & 21:27:43 & 53732.898816 & F814W & 1220  \\
 2005-12-28 & 21:50:19 & 53732.914510 & F814W & 1220  \\
 2005-12-30 & 13:28:07 & 53734.565643 & F475W & 1200  \\
 2005-12-30 & 13:50:23 & 53734.581106 & F475W & 1200  \\
 2005-12-30 & 15:01:31 & 53734.630619 & F814W & 1220  \\
 2005-12-30 & 15:24:07 & 53734.646314 & F814W & 1220  \\
 2005-12-31 & 21:25:43 & 53735.897307 & F475W & 1200  \\
 2005-12-31 & 21:47:59 & 53735.912770 & F475W & 1200  \\
 2005-12-31 & 22:59:50 & 53735.962781 & F814W & 1220  \\
 2005-12-31 & 23:22:26 & 53735.978476 & F814W & 1220  \\
 2006-01-02 & 19:47:55 & 53737.829384 & F475W & 1200  \\
 2006-01-02 & 20:10:11 & 53737.844847 & F475W & 1200  \\
 2006-01-02 & 21:21:19 & 53737.894361 & F814W & 1220  \\
 2006-01-02 & 21:43:55 & 53737.910055 & F814W & 1220  \\
 2006-01-05 & 18:07:35 & 53740.759696 & F475W & 1200  \\
 2006-01-05 & 18:29:51 & 53740.775159 & F475W & 1200  \\
 2006-01-05 & 19:42:00 & 53740.825379 & F814W & 1220  \\
 2006-01-05 & 20:04:36 & 53740.841073 & F814W & 1220  \\
 2006-01-08 & 13:15:26 & 53743.556798 & F475W & 1200  \\
 2006-01-08 & 13:37:42 & 53743.572261 & F475W & 1200  \\
 2006-01-08 & 14:49:42 & 53743.622376 & F814W & 1220  \\
 2006-01-08 & 15:12:18 & 53743.638071 & F814W & 1220  \\
\hline
\end{tabular}
\end{minipage}
\end{table}


 Leo~A has been observed as part of the LCID\footnote{Local Cosmology from Isolated
 Dwarfs: http://www.iac.es/project/LCID/.} project, which aims at constraining
 the star formation histories (SFH) of a sample of isolated LG galaxies based on
 deep color-magnitude diagrams (CMD) and time series photometry obtained with
 the Advanced Camera for Surveys onboard the {\it Hubble Space Telescope}
 ({\it HST}). The global SFH
 of Leo~A has been presented in \citet{col07}, while the analysis of population
 gradients and spatially resolved SFH will appear in a forthcoming paper
 (S.\ Hidalgo et al. 2013, in preparation). Here we present the search for
 short-period variable stars, in particular Cepheids, for which these data are
 particularly well suited.

 The paper is organized as follows: the data acquisition, reduction, and
 calibration are summarized in Section~\ref{sec:obse}. We describe the
 identification of variable stars and their spatial distribution in
 Section~\ref{sec:variables}. Sections~\ref{sec:rrl} to \ref{sec:eb} present the
 properties of the variable stars and candidate variables detected. The distance
 estimates based on various tracers and methods are presented in
 Section~\ref{sec:dist}, and we discuss the implications of our results in
 Section~\ref{sec:discuss}. In Section~\ref{sec:conclusions} we summarize our
 results and present our conclusions.

\section{Observations and Data Reduction}\label{sec:obse}

 The observations of Leo~A were carried out with the ACS onboard the {\it HST}
 between 2005 December 26 and 2006 January 8. The observed
 field was slightly offset from the center of the galaxy, as shown in
 Figure~\ref{fig:loc}, to cover a longer radial baseline and study possible
 stellar population gradients.
 Sixteen {\it HST} orbits were devoted to this galaxy in two bands (F475W and
 F814W). Each orbit was split into two $\sim$1200 second exposures of a
 given band. This resulted in 16 images per band, for a total
 exposure time of 19,200 and 19,520s in F475W and F814W, respectively. The
 complete observing log is presented in Table~\ref{tab:tab1}.

 We note that the temporal sampling for this galaxy is not as optimal as was the
 case in the previous papers of this series. In our study of variable stars in
 Cetus and Tucana \citep[hereinafter Paper I]{ber09} and IC\,1613
 \citep[hereinafter Paper II]{ber10}, we had 25, 32, and 24 datapoints per band,
 respectively, compared to only 16 here. In addition, the observing strategy for
 these three galaxies provided alternating observations in F475W and F814W, and
 therefore observations in a given filter that were well separated in time. For
 Leo~A, the paired observations in a given band are only 22 minutes apart, thus
 limiting their utility. While it did not affect our ability to detect candidate
 variables, some light curves have gaps which could limit the accuracy of the
 measured mean magnitudes (see Section~\ref{sec:cep}).

\begin{table*}
 \centering
 \begin{minipage}{200mm}
  \caption{Photometry of the Variable and Candidate Variable Stars. The complete table is available from the online edition.\label{tab:tab2}}
  \begin{tabular}{@{}ccccccccccccc@{}}
  \hline
MHJD\footnote{Modified HJD of mid-exposure: HJD$-$2,400,000.} & F475W & $\sigma_{475}$ & B$^c$ & $\sigma_B$ &
MHJD\footnote{Midpoint of F475W and F814W MHJD.} & V\footnote{Transformed from the F475W and F814W magnitudes using the equations of \citetalias{ber09}.} & $\sigma_V$ &
MHJD & F814W & $\sigma_{814}$ & I$^c$ & $\sigma_I$ \\
 \hline
\multicolumn{13}{c}{V001} \\ \hline
 53730.369406 & 25.296 & 0.030 & 25.442 & 0.030 &  53730.401888 & 24.978 & 0.030 &  53730.434371 & 24.370 & 0.061 & 24.356 & 0.061 \\
 53730.384869 & 25.284 & 0.056 & 25.418 & 0.056 &  53730.417468 & 24.986 & 0.056 &  53730.450066 & 24.416 & 0.036 & 24.400 & 0.036 \\
 53730.835645 & 25.181 & 0.032 & 25.311 & 0.032 &  53730.868127 & 24.889 & 0.032 &  53730.900610 & 24.330 & 0.034 & 24.314 & 0.034 \\
 53730.851108 & 25.239 & 0.048 & 25.385 & 0.048 &  53730.883706 & 24.921 & 0.048 &  53730.916305 & 24.314 & 0.033 & 24.300 & 0.033 \\
 53732.833851 & 25.107 & 0.031 & 25.238 & 0.031 &  53732.866334 & 24.813 & 0.031 &  53732.898816 & 24.252 & 0.036 & 24.236 & 0.036 \\
\hline
\end{tabular}
\end{minipage}
\end{table*}


 The DAOPHOT/ALLFRAME suite of programs \citep{ste94} was used to obtain the
 instrumental photometry of the stars on the individual, non-drizzled images
 provided by the {\it HST} pipeline.
 Additionally, we used the pixel area maps and data quality masks to correct for
 the pixel area variation and to flag bad
 pixels. Standard calibration was carried out as described in \citet{sir05}.
 We refer the reader to \citet{mon10} for a detailed description of the
 data reduction and calibration. The final CMD is presented in
 Figure~\ref{fig:cmd}, where the $(F475W+F814W)/2 \sim V$ filter combination was
 chosen for the ordinate axis so that the horizontal-branch (HB) appears
 approximately horizontal.

 The light-curves of the variable stars were then converted to Johnson-Cousins
 BVI magnitudes using the transformations given \citetalias{ber09}. As we have
 shown \citetalias{ber10} from the comparison of the light curves of Cepheids in
 common between our HST survey and OGLE\footnote{Optical Gravitational Lensing
 Experiment: http://ogle.astrouw.edu.pl/}, these transformations can be safely
 used to analyse the properties (e.g.\ magnitude and amplitude) of the variable
 stars.

\section{Variable stars}\label{sec:variables}

   \subsection{Identification and period search}\label{sec:period}

\begin{figure}
 \includegraphics[width=7cm]{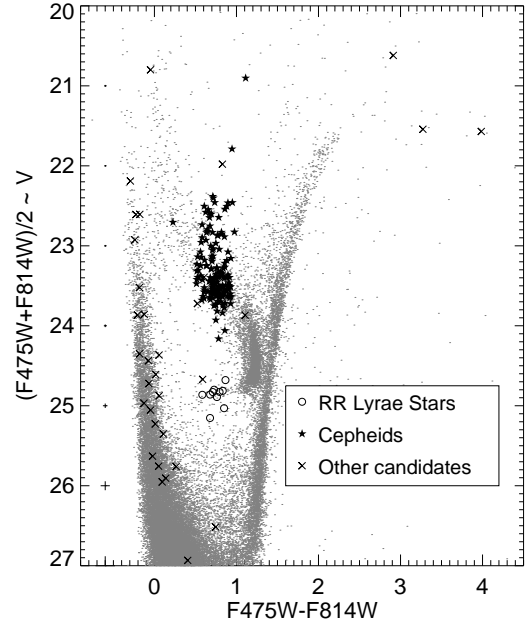}
 \caption{Color-magnitude diagram of Leo~A from our ACS data. The variables are
 overplotted as labeled in the inset.}
 \label{fig:cmd}
\end{figure}

 The candidate variables were extracted from the photometric catalogs using the
 variability index of \citet{wel93}; this process yielded $\sim$700 candidates.
 A preliminary check of the light-curve and position on the CMD, together with
 a careful inspection of the stacked image, allowed us to discard false
 detections due to cosmic-ray hits, chip defects or stars located under the
 wings of bright stars. Our final catalogue contains 166 bona fide variables and
 33 candidates for which we could not determine a period (e.g.\ main-sequence
 and red giant branch [RGB] variables, eclipsing binaries).
 The variables and candidates are shown overplotted on the CMD in
 Figure~\ref{fig:cmd} using their intensity-averaged mean magnitudes. The
 individual F475W and F814W measurements for all the variables and candidates,
 as well as the transformed BVI magnitudes, are listed in Table~\ref{tab:tab2}.

 The period search was performed through Fourier analysis \citep{hor86} taking
 into account the information from both bands simultaneously, then refined
 interactively by modifying the period until a tighter light curve was obtained.
 For each variable, datapoints with error bars larger than 3-$\sigma$ above the
 mean error bar size were rejected through sigma clipping with five iterations.
 The periods are given with three significant figures only to reflect the
 sparse sampling of the light curves. Note that for the same reason, some of the
 periods may be aliases of the true periods.

\begin{figure*}
 \includegraphics[width=12cm]{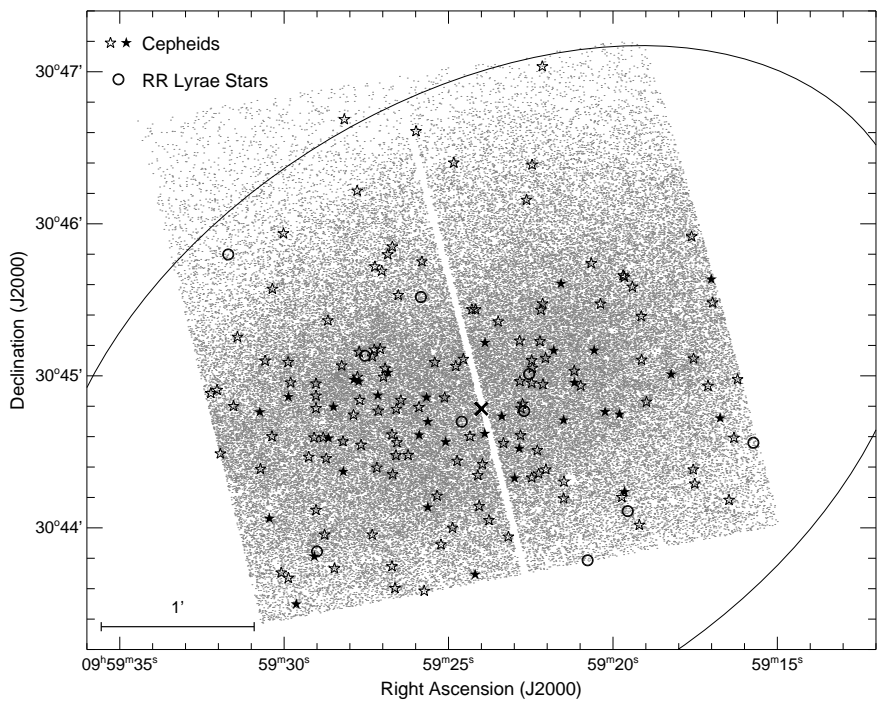}
 \caption{Spatial distribution of stars in our observed field, where the
 variables are overplotted as labeled in the inset. The cross and ellipse
 represent the center and Holmberg radius of Leo~A
 \citep[3.5$\arcmin$;][]{van04}.}
 \label{fig:distrib}
\end{figure*}

 The classification of the candidates was based on their light curve morphology
 and position in the CMD. We found a large number of Cepheids but very few
 RR~Lyrae stars (156 and 10, respectively), in very good agreement with the
 study of \citetalias{dol02}. The small number of RR~Lyrae stars -- which is not
 due to incompleteness, our photometry being near 100 percent complete at the
 magnitude of the HB -- is not surprising considering the SFH of
 Leo~A \citep[S.\ Hidalgo et al.\ 2013, in preparation]{col07}, which shows a
 very low star formation rate for the earliest epochs.
 To obtain the amplitudes and intensity-averaged magnitudes of the variables, we
 fitted the light curves with a set of templates based on the set of
 \citet[][see \citetalias{ber09}]{lay99}.

 Table~\ref{tab:tab3} summarizes the properties of the bona fide variable stars
 in our final catalogue. The first two columns give the identification number
 and type of variability, while the next two list the equatorial coordinates
 (J2000.0). Columns (5) and (6) give the period and logarithm of the period, in
 days. The intensity-averaged mean magnitudes $\langle F475W \rangle$ and
 $\langle F814W \rangle$, and color $\langle F475W \rangle -
 \langle F814W \rangle$ are given in columns (7), (9), and (11). The amplitudes
 in the F475W and F814W bands measured from the template fits are listed in
 columns (8) and (10). The last six columns alternately list the
 intensity-averaged mean magnitudes and amplitudes in the Johnson B, V, and I
 bands. For the candidate variables, we only list the coordinates and
 approximate magnitudes in Table~\ref{tab:tab4}.

   \subsection{Spatial Distribution}\label{sec:distrib}

 Figure~\ref{fig:distrib} presents the spatial distribution of stars in our
 field-of-view (FOV), where the RR~Lyrae stars (open circles) and Cepheids (open
 and filled stars) are shown. The cross and ellipse represent the center and
 Holmberg radius of Leo~A \citep[3.5$\arcmin$, from][]{van04}. Despite their
 small number, the RR~Lyrae stars seem to be roughly uniformly distributed over
 the field. On the other hand, the Cepheids are mostly located close to the
 center of the galaxy, where the main sequence stars are strongly concentrated
 \citepalias[S.~Hidalgo et al.\ 2013, in preparation]{dol02}. This shows that
 star formation in the past $\sim$1~Gyr was confined to the central region of
 the galaxy.

 We further separated the Cepheids into a bright sample and a faint sample,
 with a division at W$_{814}$=22 (see Section~\ref{sec:cep}). These samples are
 shown as filled and open stars, respectively. Interestingly, the Cepheids of
 the bright sample are even more concentrated close to the center of the galaxy
 than the whole sample, all of them being located in the lower half of our FOV.
 These gradients indicate that the star forming region has been shrinking with
 time \citep[see e.g.][]{hid09}, at least in the past Gyr or so.
 This result is in excellent agreement with the SFH as a function of
 galactocentric radius of S.\ Hidalgo et al.\ (2013, in preparation).

\begin{figure}
 \includegraphics[width=7cm]{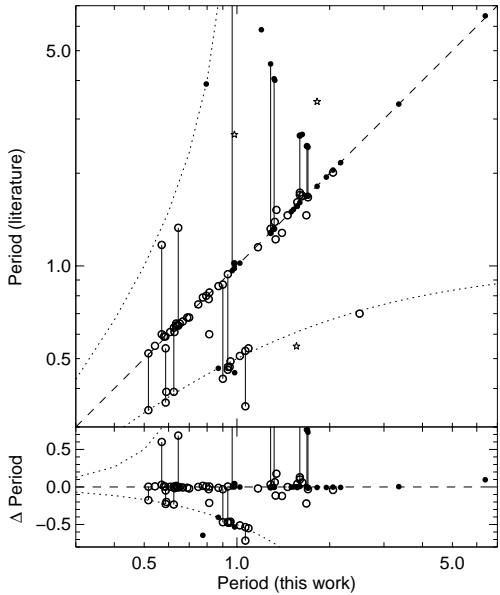}
 \caption{Period difference for the variable stars in common between this and
 previous works: \citetalias{dol02} (open circles), \citetalias{goe07} (filled
 circles), and \citetalias{hoe94} (open stars).
 When two periods were listed for a given star, these are joined by vertical
 lines. The dashed line is the identity line, and the dotted lines represent the
 one-day aliases of the best period.}
 \label{fig:perdiff}
\end{figure}

   \subsection{Comparison with previous catalogues}\label{sec:comp}

 The search for variable stars in Leo~A has been carried out by several authors
 prior to this work. We cross-matched the catalogues from the literature with
 ours to identify the stars in common and compare the measured properties. These
 are listed in Table~\ref{tab:tab5}. We give their index, type, and period
 measured from our data, together with their identifications and periods from
 previous catalogues\footnote{The J2000.0 coordinates of all the variable star
 candidates from \citet{san86}, \citet{hoe94}, and \citetalias{dol02}, which are
 not listed in the original publications, are available from the first author on
 request.}.

 Variable stars in Leo~A were first mentioned by \citet{san86}. While the data
 were not sufficient for a detailed study, Sandage found eight candidates by
 blinking three plate pairs. However, only the four best candidates are labeled
 on the finding chart, and no information on periods or magnitudes are given.
 By comparing his finding charts with ours, we found the following:
 his ``definite variable'' (RA=09:59:19.52, Dec=+30:44:13.4) is a very bright MS
 star; one candidate (RA=09:59:28.88, Dec=+30:45:42.6) is a blend of two
 relatively bright RGB stars, and another (RA=09:59:24.70, Dec=+30:44:48.1) is
 close to the tip of the RGB (TRGB); none of these appear to be variable in our
 data, possibly due to having periods longer than our observational baseline.
 The fourth candidate is out of our FOV, but corresponds to the D02 Cepheid
 C2-V67.

\begin{figure}
 \includegraphics[width=8cm]{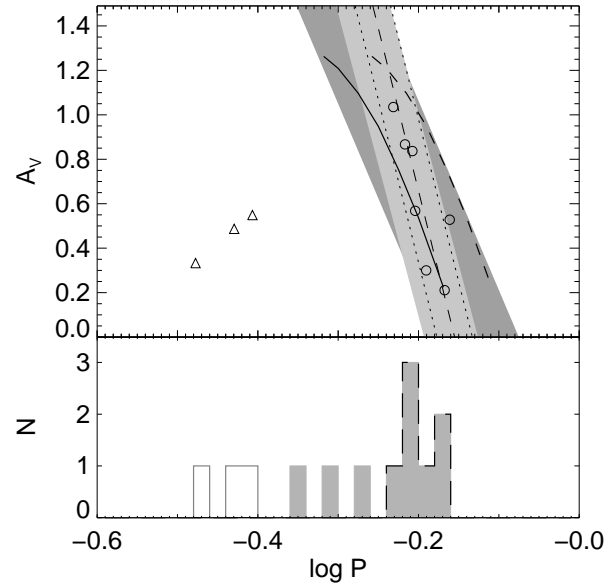}
 \caption{Top: Period-amplitude diagram for the RR Lyrae stars. Circles and
 triangles represent RR$ab$ and RR$c$, respectively. The thin dashed line is a
 fit to the period-amplitude of the RR$ab$; the dotted lines show the
 1.5-$\sigma$ limits. The light and dark grayed areas represent the
 $\pm$1.5-$\sigma$ limits of Cetus and Tucana, respectively, from
 \citetalias{ber09}. The solid and dashed curved lines show the loci of OoI and
 OoII globular clusters from \citet{cac05}.
 Bottom: Period histogram for the RR Lyrae stars of the top panel. RR$ab$ and
 RR$c$ are shown as the black dashed and solid gray lines, respectively, while
 the filled histogram represent the fundamentalized RR Lyrae stars.}
 \label{fig:peramp}
\end{figure}

 The first actual search for variable stars was carried out by
 \citet[][hereinafter H94]{hoe94} using data from a variety of telescopes and
 instruments, from which they discovered 14 suspected variables (including five
 Cepheids: V5, V8, V9, V10, and V14 in their catalogue). Out of these, V1, V2,
 and V14 are located outside of our FOV. The stars V3--V6, V8, and V12 are not
 variables in our data, and V11 is a long-period variable (LPV). We classify the
 remaining four candidates (V7, V9, V10, V13) as Cepheids, but the periods from
 \citetalias{hoe94} cannot phase our data.

 More recently, \citetalias{dol02} analysed 28 images (23 in V, five in R) taken
 with the WIYN 3.5-m telescope, and found 92 candidates variables (84 Cepheids
 and 8 RR~Lyrae stars). Our catalogue contains 49 of the 50 variables in our FOV.
 Their C2-V33 appears to be a (non-variable) red clump (RC) star in our data. As
 shown in Figure~\ref{fig:perdiff}, for most of the stars in common we find a
 very good agreement between their reported periods and ours. For the few
 variables with discrepant periods, we checked that their value did not give a
 better light curve.
 The dotted lines in Figure~\ref{fig:perdiff} show that most of the inadequate
 periods of \citetalias{dol02} are due to the one-day aliasing usually affecting
 ground-based data.

 The most recent work on variable stars in this galaxy is by the Munich group
 \citep[hereinafter G07]{sni06,goe06,goe07} using small telescopes with the
 aim of studying the properties of bright variables in a sample of six northern
 dwarf irregular galaxies, including Leo~A. The work of \citetalias{goe07}
 focused on Cepheids while \citet{sni06} concentrated on LPVs, although some of
 the stars with unsecure classification appear in both. Given the relatively
 short time baseline of our observations preventing us from detecting long
 periodicities, most of the previously known stars in our catalogue are in
 common with \citetalias{goe07}.
 In column 7 of Table~\ref{tab:tab5} we use the star indices of
 \citetalias[where LS, LM, and LL refer to short, medium, and long period
 variables, with cut-offs at 1 and 10 days]{goe07}, but also list the indices
 from \citet[starting with V-]{goe06} and from \citet[starting with
 LeoA-]{sni06}, when available.

\begin{figure}
 \includegraphics[width=9.0cm]{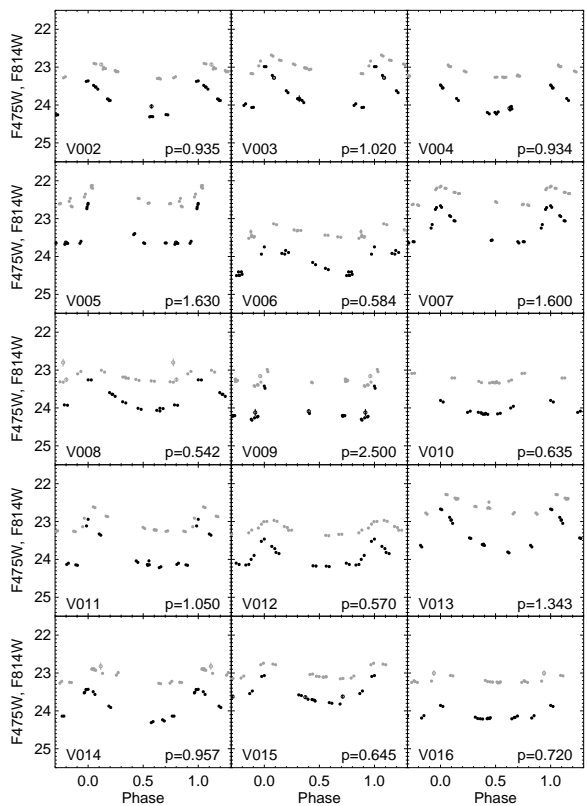}
 \caption{Sample light curves of the Cepheids stars in the $F475W$ (black) and
 $F814W$ (gray) bands, phased with the period in days given in the lower right
 corner of each panel. Open symbols show the bad data points, i.e.\ with errors
 larger than 3-$\sigma$ above the mean error of a given star, which were not
 used in the calculation of the period and mean magnitudes. For a few very
 bright Cepheids, the light curves have been shifted downward by 1 or 2 mag, as
 shown by an arrow in the top left corner of a panel.
 The complete sample of light curves is available in the on-line version.}
 \label{fig:cep_lc}
\end{figure}

The catalogue of \citetalias{goe07} contains 131 candidate variables with
 R$<$23.5 and P$<$130 days. Of these, 33 are matches to Cepheids in our
 catalogue, including 19 with periods in good agreement with ours (see
 Figure~\ref{fig:perdiff}). The remaining ones are either outside of our FOV
 (45), or not variable in our data (53). We note, however, that all their
 candidates flagged as `best' and `good' and located in our FOV are indeed
 Cepheids.

\begin{figure}
 \includegraphics[width=8cm]{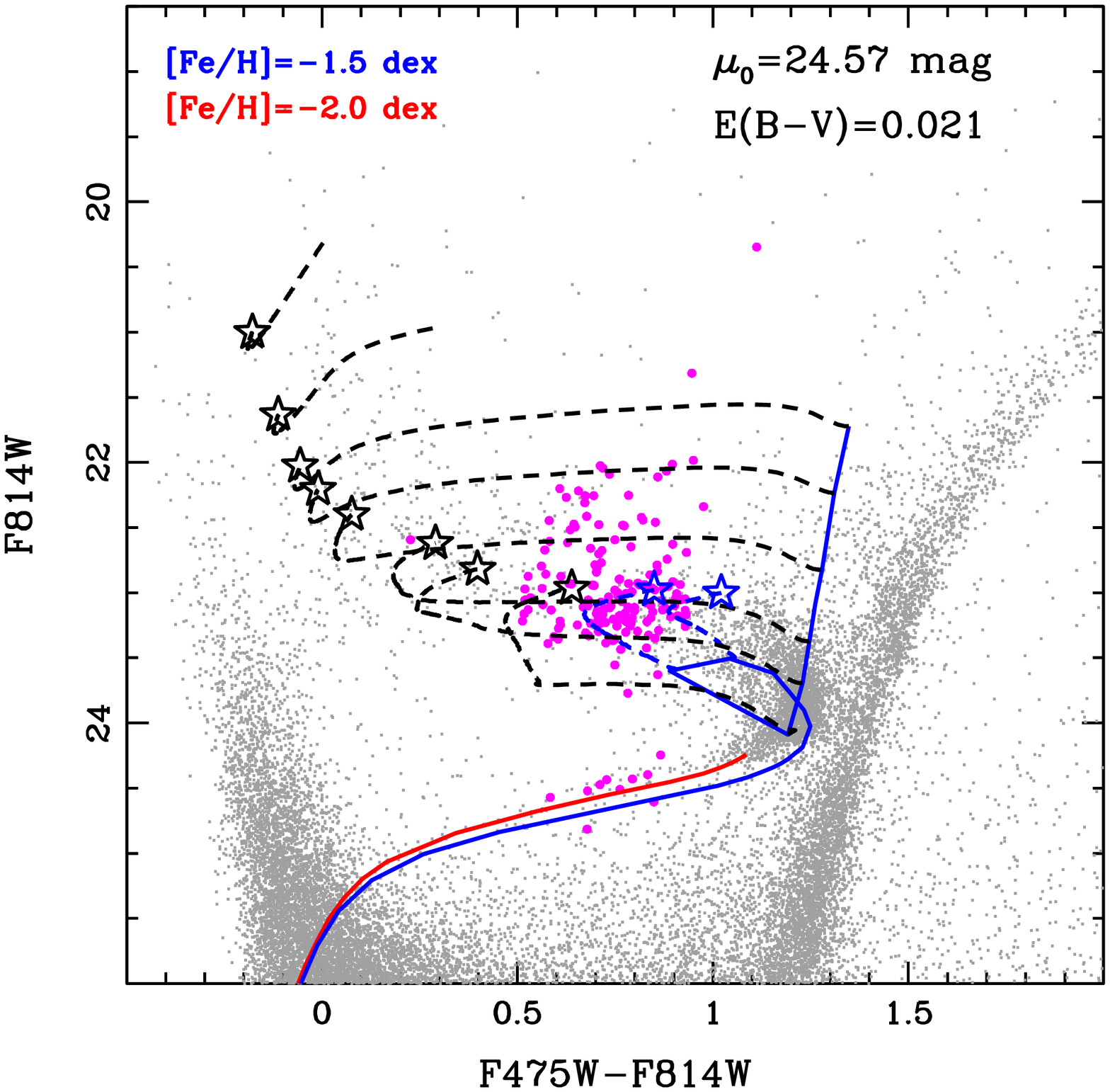}
 \caption{CMD showing the observed variable (magenta filled circles) and
 non-variable (gray dots) stars, where theoretical predictions from the BaSTI
 library \citep{pie04} are overplotted. The thick blue line shows the ZAHeB locus
 for [Fe/H]=$-$1.5. Evolutionary tracks for both anomalous (blue dashed lines,
 M = 2.0 and 2.14~M$_{\sun}$) and classical (black dashed lines, M = 2.2, 2.4,
 2.6, 3.0, 3.5, 4.0, 5.0, and 7.0~M$_{\sun}$) Cepheids are shown for the same
 metallicity. The stellar tracks extend from the ZAHeB to the central helium
 exhaustion, represented by the open star symbols. The thick red line shows the
 ZAHeB locus for [Fe/H]=$-$2.0, which provides a much better fit to the old
 stellar populations of the HB (see Section~\ref{sec:discuss1}).}
 \label{fig:tracks}
\end{figure}

\section{RR Lyrae Stars}\label{sec:rrl}

 Ten of the variable stars in our sample are located on the HB
 and were classified as RR~Lyrae stars. From their light curves and periods, we
 further classify them as 7 fundamental-mode (RR$ab$) and 3 overtone (RR$c$)
 pulsators. We find mean periods of 0.636 and 0.366 day for the RR$ab$ and RR$c$,
 respectively, and a fraction of overtone pulsators of
 $f_c=N_c/(N_{ab}+N_c)=0.3$. These numbers place Leo~A in the
 \citet{oos39} type II group, whereas the vast majority of Local Group dwarf
 galaxies have properties intermediate between those of the type I and type II
 groups \citep[OoI and OoII, respectively; see][]{smi09,cat09}. We note, however,
 that the reliability of these numbers may be affected by the small number of
 RR~Lyrae stars and suboptimal light curves.

 In fact, this classification is not as clear from the location of these
 variables in the period-amplitude diagram.
 Figure~\ref{fig:peramp} presents the period-amplitude diagram ({\it top}) and
 the period distribution ({\it bottom}) of the RR~Lyrae stars. In the top panel,
 RR$ab$ and RR$c$ are shown as open circles and open triangles, respectively.
 The solid and dashed curved lines show the loci of OoI and OoII globular
 clusters from \citet{cac05}. The low-amplitude RR$ab$ stars seem to lie on the
 OoI locus, while the ones with higher amplitude are located between the two
 groups, suggesting a Oosterhoff-intermediate classification instead. The
 relatively long mean period for the RR$ab$ stars may thus be due to the lack of
 high-amplitude RR$ab$ stars.

 The shaded areas in the top panel of Figure~\ref{fig:peramp} represent the
 location of RR$ab$ stars in Cetus and Tucana
 from \citetalias{ber09}. The figure shows that the distribution of the RR$ab$
 of Leo~A in period-amplitude space, forming a tight and almost vertical
 sequence, is very similar to that of Cetus. On the other hand, in
 \citetalias{ber10} we found that the RR$ab$ in IC\,1613 were more similar to
 those in Tucana. Interestingly, both Cetus and Leo~A seem to have had a
 significantly lower SFR at very early epochs than Tucana and IC\,1613, which
 may partly explain the difference.

 The very low number of RR~Lyrae stars is a remarkable result given the
 high stellar density in our FOV and completeness at the magnitude of
 the HB ($\ga$98 percent). For comparison, we found 90 or more RR~Lyrae stars in
 roughly similar areas in the other LCID galaxies \citepalias{ber09,ber10}.
 To make this comparison more meaningful, for each of these galaxies we
 calculated the number of RR~Lyrae stars relative to the number of RGB stars
 within the 2.5 magnitudes below the TRGB. We find 0.01, 0.09, 0.58,
 and 0.81 for Leo~A, IC\,1613, Cetus, and Tucana, respectively.
 Assuming RGB and RR~Lyrae stars are older than 1.5 and 10~Gyr, respectively,
 this ratio gives a measure of the fraction of very old stellar population with
 respect to the intermediate-age and old stellar populations of the galaxy.
 The extremely low value found for Leo~A is in very good agreement with the
 results of the SFH analysis. \citet{col07} and S.\ Hidalgo et al.\ (2013, in
 preparation) found very little, if any, star formation for ages older than
 10~Gyr, and showed that 90 percent of the star formation had occurred more
 recently than 8 Gyr ago.

\section{Cepheids}\label{sec:cep}

 Most of the variable stars in our sample are found in the classical instability
 strip above the HB. We present their light curves in Figure~\ref{fig:cep_lc}.
 Recent work has shown that at the metallicity of the young population of Leo~A
 ([Fe/H]~$\sim-$1.5), CC and AC can occupy a similar region in the CMD
 \citep{fio12}. CC are young, relatively massive stars ($\ga$2.2~M$_{\sun}$) in
 the phase of core helium burning. AC, on the other hand, are older and less
 massive stars on the HB phase of evolution, burning helium in the core in
 partially degenerate conditions after undergoing the so-called helium flash at
 the TRGB \citep[e.g.][]{dem75}.

 In Figure~\ref{fig:tracks} we show a comparison of the observed location of the
 Cepheids in the CMD with the theoretical predictions for both CC and AC with
 masses 2.0--7.0~M$_{\sun}$. The models are based on the BaSTI stellar evolution
 library \citep{pie04}, and were shifted to the distance and foreground
 reddening of Leo~A. The thick blue line shows the zero-age helium-burning
 (ZAHeB) locus, which represents the ignition of helium in the core in either
 degenerate (M~$\le$~2.14~M$_{\sun}$ at [Fe/H]~$=-$1.5) or non-degenerate
 (M~$>$~2.14~M$_{\sun}$) conditions. The black and blue dashed lines show the
 theoretical tracks for CC and AC, respectively. They extend from the ZAHeB to
 the open star symbols representing exhaustion of helium in the core.
 The Figure shows that while all the Cepheids brighter than F814W$\sim$23 are
 most likely CC, the fainter ones could be a combination of both CC and AC. Note
 that this result does not change for significantly lower metallicity
 (i.e.\ [Fe/H]=$-$1.8). On the other hand, for [Fe/H]=$-$1.3 and higher, the AC
 tracks do not enter the instability strip, leading to a pure CC population.

\begin{figure}
 \includegraphics[width=8cm]{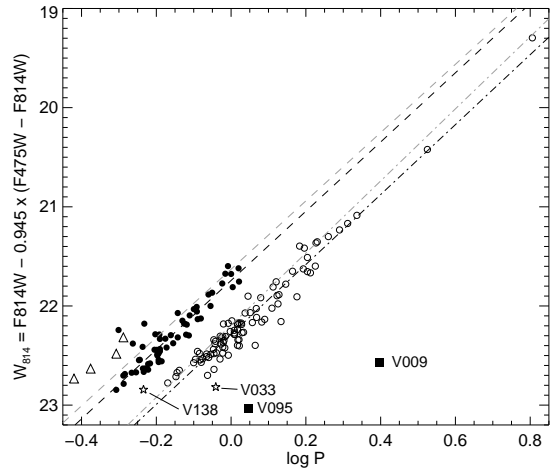}
 \caption{Period-Wesenheit diagram for Cepheids in Leo~A. Open circles, filled
 circles, and open triangles represent fundamental, first overtone, and second
 overtone pulsators, respectively. A few outliers are labeled and discussed in
 Section~\ref{sec:cep}. The black dash-dotted and dashed lines are
 linear fits to the fundamental and FO Cepheids. The gray lines are the same
 fits to the IC1613 Cepheids \citepalias{ber10}, shifted to the distance of
 Leo~A.}
 \label{fig:wesen}
\end{figure}

\begin{figure*}
 \includegraphics[width=17.7cm]{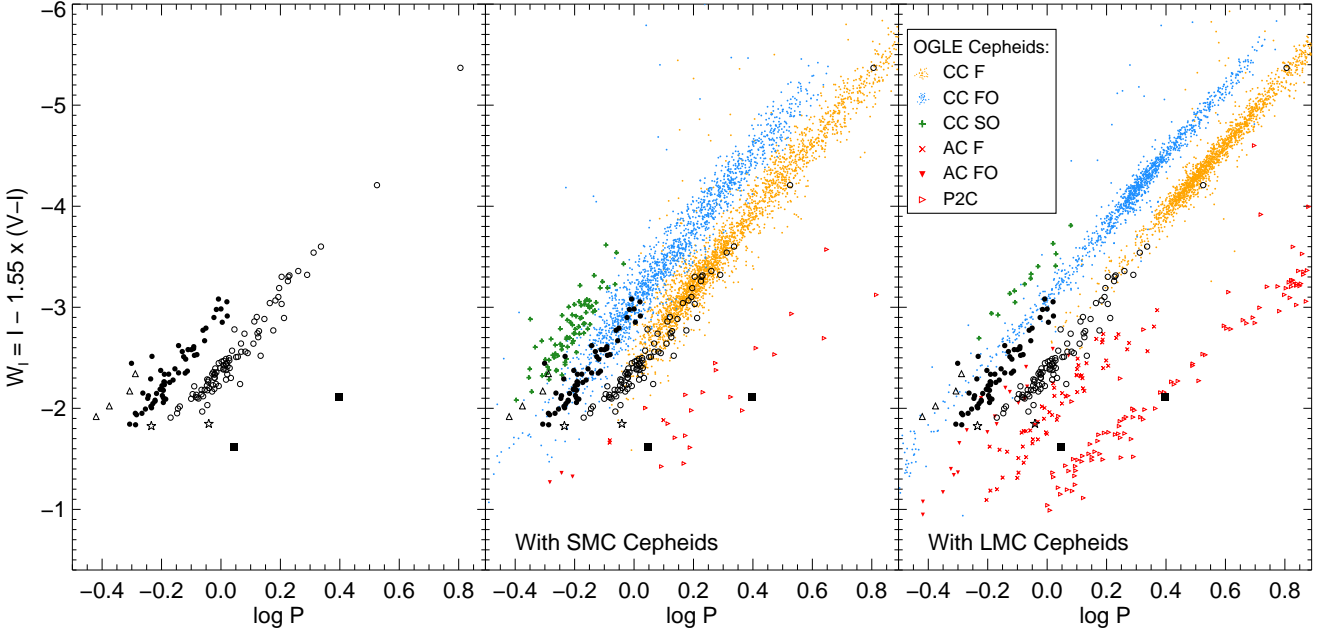}
 \caption{Period-luminosity diagram for Cepheids in Leo~A (black symbols as in
 Figure~\ref{fig:wesen}). The left panel shows the Leo A data alone, while in
 the middle and right panels the Cepheids in the SMC and LMC, respectively, are
 also shown. In the inset, CC, AC, and P2C stand for classical, anomalous, and
 population II Cepheids; F, FO, and SO indicate fundamental, first-, and
 second-overtone mode pulsations.}
 \label{fig:pl}
\end{figure*}

 Since the various types of Cepheids have different magnitudes at a given
 period, their locations in the PL diagrams can be used to differentiate their
 types \citep[e.g.,][]{fio06,mon12}. Figure~\ref{fig:wesen} shows the
 period-luminosity diagram of all the Cepheids in Leo A. Here we used the
 reddening-free Wesenheit magnitude
\begin{displaymath}
 W_{814} = F814W - 0.945 \times (F475W-F814W)
\end{displaymath}
 in order to reduce the scatter due to interstellar extinction and the intrinsic
 dispersion due to the finite width of the instability strip. Despite the
 sub-optimal sampling of our light curves, the sequences of fundamental
 (F, open circles) and first-overtone (FO, filled circles) mode Cepheids are
 very well defined. The black dash-dotted and dashed lines represent linear
 fits to the F and FO Cepheids, respectively. The grey lines represent similar
 fits to the IC\,1613 variables of \citetalias{ber10}, shifted to the distance
 of Leo~A, and are discussed in Section~\ref{sec:Cepdist}. The four Cepheids
 shown as open triangles, $\sim$0.4 brighter than the FO relation, have
 sinusoidal light curves with very low amplitudes (A$_{F475W} \la$~0.2~mag, see
 Figure~\ref{fig:cep_lc}), and are located along the blue edge of the
 instability strip. We therefore classify them as second-overtone (SO)
 mode Cepheids. A few outliers are labeled and discussed below.
 From this plot, we classify 91 and 58 stars as F and FO Cepheids, respectively.

 Figure~\ref{fig:pl} shows the Leo~A Cepheids in the period-Wesenheit ($P-W_I$)
 diagram, together with the OGLE-III Cepheids of the Small and
 Large Magellanic Clouds \citep[LMC and SMC,
 respectively;][]{sos08a,sos08b,sos10a,sos10b} in the middle and
 right panels. In all three panels, the Leo~A Cepheids are shown with the same
 symbols as in Figure~\ref{fig:wesen}, and are shown by themselves in the left
 panel for clarity. The Cepheids of Leo~A were shifted according to the TRGB
 distance determined in Section~\ref{sec:distTRGB}, i.e.\ (m$-$M)$_0=$24.57.
 The LMC/SMC apparent magnitudes were converted to absolute magnitudes assuming
 a distance modulus of 18.515$\pm$0.085 to the LMC \citep{cle03}, and a distance
 offset of 0.51 of the SMC relative to the LMC \citep{uda99c}. The middle panel
 shows that the bulk of Leo~A Cepheids are in relatively good agreement with the
 SMC Cepheids -- even though the significant spread of the SMC Cepheids
 complicates the comparison -- and extend the F, FO, and SO PL relations to
 shorter periods and fainter magnitudes. This supports their identification as
 CC.

 We note, however, that at the shortest period the FO Cepheids of Leo~A and the
 SMC seem to be slightly offset, the former appearing $\sim$0.1~mag fainter than
 the latter at log\,P~$\sim-$0.25. This is more obvious in the right panel,
 which shows that they are fainter by $\sim$0.2--0.3~mag than the LMC classical
 Cepheids at a given period. Part of these offsets may be due to uncertainties
 in the distance moduli used. On the other hand, we also find small differences in
 the slope of the PL relations calculated by linear regression over the period
 ranges $-0.2<{\rm log\, P}<1.0$ and $-0.4<{\rm log\, P}<0.8$ for the F and FO
 Cepheids, respectively. For the F and FO Cepheids of Leo~A, we find slopes of
 $-$3.538$\pm$0.074 and $-$3.58$\pm$0.15, respectively, comparable to those of
 the SMC ($-$3.522$\pm$0.013 and $-$3.639$\pm$0.020), but mildly steeper than
 those of the LMC ($-$3.332$\pm$0.012 and $-$3.439$\pm$0.011). The possible
 origin of the offset is further discussed in Section~\ref{sec:discuss}.

 Despite this offset, most of the Leo~A Cepheids lie closer to the sequences of
 F and FO classical Cepheids of the Magellanic Clouds than to the other types of
 Cepheids (AC and P2C). The Cepheids that were flagged as
 outliers in Figure~\ref{fig:wesen} fall squarely on the sequences of other
 types of Cepheids. In particular, the two faint Cepheids with periods longer
 than 1~day (V009 and V095, filled squares) lie on the P2C sequence of both
 Magellanic Clouds. V138, shown as an open star, lies between the F and
 FO classical Cepheids PL, but fits very well with the FO AC of the LMC.
 Some of the other Cepheids that are slightly offset from the
 classical F and FO sequences (e.g.\ V033 in Figure~\ref{fig:wesen}) may actually
 be AC. Unfortunately, our light curves are not sufficiently densely populated
 to classify the Cepheids based on light curve morphology. The follow-up
 observations of Fiorentino et al.\ (2013, in preparation), carried out with
 the Gemini telescope, may help us to confirm the presence of AC in
 Leo~A thanks to the better sampling.

\section{Other Candidate Variables}\label{sec:eb}

\begin{table*}
 \centering
 \begin{minipage}{200mm}
  \caption{Pulsational Properties of the Variable Stars. The complete table is available from the online edition.\label{tab:tab3}}
  \begin{tabular}{@{}cccclcccccccccccc@{}}
  \hline
ID & Type\footnote{$ab$ and $c$: fundamental mode and overtone RR~Lyrae star; F, FO, SO: fundamental mode, first-, and second-overtone Classical Cepheid; P2C: population II Cepheid; Cep: Cepheid of unknown type.} & 
R.A. & Decl. & Period & log P & $\langle F475W \rangle$ & A$_{475}$ & $\langle F814W \rangle$ & A$_{814}$ & $\langle F475W \rangle-$ &
$\langle B \rangle$ & A$_{B}$ & $\langle V \rangle$ & A$_{V}$ & $\langle I \rangle$ & A$_{I}$ \\
   &      & (J2000) & (J2000) & (days) & (days) &  &  &  &  & $\langle F814W \rangle$ &  &  &  &  &  & \\
 \hline
 V001 & $ab$ & 09 59 15.73 & +30 44 33.6 & 0.690  & -0.161 & 25.112 & 0.552 & 24.246 & 0.436 & 0.865 & 25.211 & 0.657 & 24.796 & 0.528 & 24.213 & 0.522  \\
 V002 &   F  & 09 59 16.21 & +30 44 58.6 & 0.935  & -0.029 & 23.941 & 0.964 & 23.100 & 0.551 & 0.841 & 24.061 & 1.065 & 23.669 & 0.780 & 23.104 & 0.477  \\
 V003 &   F  & 09 59 16.32 & +30 44 35.5 & 1.02   &  0.009 & 23.729 & 1.186 & 22.980 & 0.567 & 0.750 & 23.843 & 1.321 & 23.473 & 0.956 & 22.965 & 0.548  \\
 V004 &   F  & 09 59 16.47 & +30 44 11.0 & 0.934  & -0.030 & 23.572 & 1.549 & 23.039 & 0.621 & 0.533 & 23.682 & 1.678 & 23.411 & 1.199 & 23.016 & 0.657  \\
 V005 &   F  & 09 59 16.74 & +30 44 43.3 & 1.63   &  0.212 & 23.263 & 1.382 & 22.442 & 0.459 & 0.821 & 23.359 & 1.615 & 22.918 & 1.203 & 22.431 & 0.475  \\
\hline
\end{tabular}
\end{minipage}
\end{table*}


 In addition to the classical instability strip variables presented above, we
 detected another 33 variable candidates (VC) throughout the CMD, for which we
 could not determine the period because of inadequate temporal sampling or
 insufficient number of datapoints. They are shown as crosses in the CMD in
 Figure~\ref{fig:cmd}. The majority of these candidate variables are located on
 the MS or close to the TRGB, and are therefore most likely eclipsing binaries
 or LPV; two of these (VC11 and VC32) are possible $\delta$-Scuti stars. Their
 coordinates and approximate magnitudes are provided in Table~\ref{tab:tab4}.
 They are labeled as main-sequences variables (MSV), classical instability strip
 variables (ISV), or LPV based on their location on the CMD.

\section{Distance to Leo A}\label{sec:dist}

\subsection{TRGB Luminosity}\label{sec:distTRGB}

 In the previous articles of this series \citepalias{ber09,ber10}, we used
 exclusively the properties of the variable stars to obtain the distance to
 the host galaxy. However, the rather sparse sampling of our light curves,
 fainter than expected Cepheids, and very low number of RR~Lyrae stars in Leo~A
 lead to somewhat uncertain measurements (see below). Therefore, we also include
 measurement of the distance using the TRGB method
 \citep[e.g.][]{dac90,lee93,sal98}, for which the dependence on age and
 metallicity is well understood. In particular, \citet{sal12} has shown that at
 the low metallicity of Leo~A, the luminosity of the TRGB is not affected by age
 (at least for ages larger than $\sim$4~Gyr old), metallicity, metal mixture,
 and helium abundance, and that the empirical and theoretical models agree
 within $\sim$0.05 mag.

\begin{figure}
 \includegraphics[width=7.8cm]{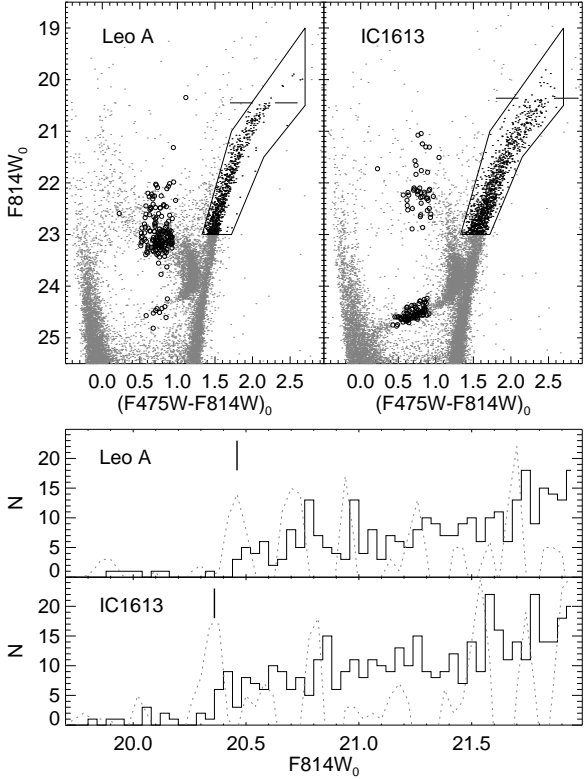}
 \caption{Top panels: CMDs of Leo~A (left) and IC\,1613 (right), showing the box
 used for the selection of RGB stars (black points). The open circles represent
 the Cepheids and RR~Lyrae stars, and the magnitude of the TRGB is marked by the
 horizontal lines near $F814W_0\sim$~20.5.
 Bottom panels: Luminosity functions of Leo~A and IC\,1613 (black histograms).
 The dotted line represents the Sobel filter response, and the location of the
 tip is shown by a vertical line.}
 \label{fig:trgb}
\end{figure}

 Figure~\ref{fig:trgb} shows the box used for the selection of RGB stars in the
 Leo~A CMD (top left panel) and corresponding luminosity function (LF, bottom
 panel). We applied the same method to IC\,1613 for comparison with the variable
 star distance from \citetalias{ber10} and to illustrate the reliability of the
 method; we show the CMD and LF in the top right and bottom panels, respectively.
 The open circles in the CMDs represent the Cepheids and RR~Lyrae stars.
 To determine the magnitude of the TRGB, we convolved the LFs with a Sobel
 kernel of the form [-1,-2,0,2,1]; the filter response function is shown as a
 dotted line in the bottom panels, and the center of the peak corresponding to
 the TRGB of each galaxy is indicated by the thick vertical line. We find
 $F814W_0=$20.46$\pm$0.12 and 20.36$\pm$0.09 for the TRGB of Leo~A and IC\,1613,
 respectively, shown as the horizontal lines in the top panels.
 For each galaxy, the errorbars were calculated by measuring the TRGB luminosity
 of 1000 synthetic CMDs containing the same number of stars in the selection box
 as the real galaxies; the quoted uncertainty is the standard deviation of these
 measurements.

 The distances were obtained from the TRGB magnitudes using the empirical
 absolute TRGB
 magnitude calibrations in the {\it HST} flight bands from \citet{riz07}. While
 these contain a color term based on either the F555W or F606W bands, which we
 do not have for Leo~A, the (F555W-F814W) color is very close to (V$-$I). We can
 therefore use the TRGB color from ground-based observations. For Leo~A and
 IC\,1613, we used (V$-$I)$_{TRGB}$=1.40$\pm$0.05 \citep{tol98} and
 1.60$\pm$0.02 \citep{riz07}, respectively. These give distance moduli of
 (m$-$M)$_{\rm Leo~A,0}=$24.57$\pm$0.13 and
 (m$-$M)$_{\rm IC\,1613,0}=$24.44$\pm$0.09. The latter value is precisely the
 same as the distance determined from Cepheids and RR~Lyrae stars in
 \citetalias{ber10}. It shows that the TRGB method used here and the variable
 stars distance methods of \citetalias{ber10} produce homogeneous and reliable
 results.
 We note that using the theoretical TRGB magnitude calibrations from
 \citet{cas13}, which takes into account the updated electron conduction
 opacities of \citet{cas07}, produces comparable values within the uncertainties
 ((m$-$M)$_{\rm Leo~A,0}=$24.58$\pm$0.13 and
 (m$-$M)$_{\rm IC\,1613,0}=$24.49$\pm$0.09).
 Our TRGB distance to Leo~A is also in good agreement with previous
 determination using this method
 \citep[(m$-$M)$_0=$~24.5$\pm$0.2;][]{tol98,sch02}.

\subsection{RR~Lyrae Stars}

 In the previous papers of the series we used both the mean magnitude of the
 RR~Lyrae stars (luminosity--metallicity relation) and the location of the
 first-overtone blue edge of the instability strip
 (period--luminosity--metallicity relation) to estimate the distance to the host
 galaxy. However, the latter method depends on having a sufficiently populated
 HB within the instability strip, and is thus not applicable here due to the
 very low number of RR~Lyrae stars found in Leo~A. Here and in the following
 section, we used the intensity-averaged mean magnitudes in the Johnson-Cousin
 bands.

 The luminosity--metallicity relation used here was derived in
 \citetalias{ber09}, and has the form:
\begin{displaymath}
 M_V = 0.866(\pm0.085) + 0.214(\pm0.047) [\rm{Fe/H}].
\end{displaymath}
 The slope was derived by \citet{cle03} from spectroscopy of a large number of
 RR~Lyrae stars in the LMC, while the zero-point was set such that the LMC
 distance modulus is (m$-$M)$_{\rm LMC,0}$=18.515$\pm$0.085 (\citealt{cle03},
 adopting [Fe/H]~$=-$1.5 for the old LMC population; \citealt{gra04}).

 We find a dereddened mean magnitude for the 10 RR~Lyrae stars of
 $\langle V \rangle_0$=24.916$\pm$0.044. The star formation and chemical
 enrichment history calculated by Hidalgo et al.\ (2013, in preparation) from
 CMD fitting gives a mean metallicity of [Fe/H]~$=-$2.0$\pm$0.1 for the old
 population. At this metallicity, the relation given above predicts a HB
 luminosity of M$_V$=0.44$\pm$0.11. The uncertainty was quantified through
 Monte-Carlo simulations. This gives a distance modulus of
 (m$-$M)$_0$=24.48$\pm$0.12 after correcting for reddening.
 We note that the distance does not change significantly if we exclude the three
 outlier RR~Lyrae stars (one bright and two faint, see Figure~\ref{fig:cmd}),
 which yields (m$-$M)$_0$=24.44$\pm$0.11. This distance is in good agreement 
 with the TRGB distance derived in Section~\ref{sec:distTRGB}.

\begin{table}
 \centering
 \begin{minipage}{85mm}
  \caption{Properties of the Candidate Variable Stars.\label{tab:tab4}}
  \begin{tabular}{@{}cccccc@{}}
\hline
ID & Type\footnote{ISV: instability strip variable; LPV: long period variable; MSV: main-sequence variable.} &   
R.A.  &  Decl.  & $F475W$ & $F814W$ \\
   &      & (J2000) & (J2000) &         &         \\
\hline
VC01 & MSV & 09:59:15.90 & +30:44:13.6 & 23.796 & 23.919 \\
VC02 & MSV & 09:59:17.69 & +30:44:54.6 & 25.232 & 25.222 \\
VC03 & MSV & 09:59:18.88 & +30:45:24.1 & 25.893 & 25.628 \\
VC04 & MSV & 09:59:18.91 & +30:45:27.6 & 24.400 & 24.473 \\
VC05 & LPV & 09:59:21.05 & +30:44:37.0 & 23.179 & 19.906 \\
VC06 & MSV & 09:59:21.80 & +30:45:19.8 & 24.688 & 24.756 \\
VC07 & ISV & 09:59:22.20 & +30:44:52.7 & 23.984 & 23.460 \\
VC08 & ISV & 09:59:22.89 & +30:45:53.1 & 24.420 & 23.314 \\
VC09 & LPV & 09:59:23.02 & +30:45:40.7 & 23.563 & 19.578 \\
VC10 & ISV & 09:59:23.37 & +30:46:01.6 & 23.696 & 23.090 \\
VC11 & MSV & 09:59:23.38 & +30:45:23.9 & 26.889 & 26.142 \\
VC12 & LPV & 09:59:25.22 & +30:45:23.7 & 22.076 & 19.164 \\
VC13 & MSV & 09:59:26.37 & +30:45:05.7 & 24.904 & 24.846 \\
VC14 & MSV & 09:59:26.44 & +30:45:01.9 & 22.518 & 22.696 \\
VC15 & MSV & 09:59:26.56 & +30:45:20.5 & 24.259 & 24.440 \\
VC16 & MSV & 09:59:26.97 & +30:44:30.1 & 25.619 & 25.642 \\
VC17 & MSV & 09:59:27.50 & +30:44:45.4 & 25.782 & 25.730 \\
VC18 & MSV & 09:59:28.03 & +30:44:17.4 & 22.496 & 22.721 \\
VC19 & MSV & 09:59:28.52 & +30:45:12.4 & 25.407 & 25.298 \\
VC20 & MSV & 09:59:28.58 & +30:45:59.0 & 26.000 & 25.905 \\
VC21 & MSV & 09:59:28.60 & +30:44:43.9 & 22.044 & 22.337 \\
VC22 & ISV & 09:59:28.70 & +30:45:46.0 & 22.394 & 21.565 \\
VC23 & MSV & 09:59:29.16 & +30:44:32.8 & 24.904 & 25.031 \\
VC24 & MSV & 09:59:29.38 & +30:44:59.3 & 23.763 & 23.970 \\
VC25 & MSV & 09:59:29.40 & +30:43:58.0 & 24.394 & 24.337 \\
VC26 & MSV & 09:59:30.21 & +30:44:41.2 & 20.776 & 20.823 \\
VC27 & MSV & 09:59:30.43 & +30:45:15.9 & 24.617 & 24.601 \\
VC28 & MSV & 09:59:30.83 & +30:44:19.5 & 22.807 & 23.045 \\
VC29 & ISV & 09:59:30.94 & +30:44:30.7 & 24.964 & 24.376 \\
VC30 & MSV & 09:59:31.26 & +30:44:41.6 & 23.431 & 23.613 \\
VC31 & MSV & 09:59:31.78 & +30:45:29.6 & 25.975 & 25.837 \\
VC32 & MSV & 09:59:32.14 & +30:45:05.1 & 27.136 & 26.730 \\
VC33 & MSV & 09:59:32.26 & +30:44:56.7 & 25.031 & 25.078 \\
\hline
\end{tabular}
\end{minipage}
\end{table}


\subsection{Cepheid Period-Luminosity(-Color) relations}\label{sec:Cepdist}

 Given the significant number of F, FO, and SO Cepheids observed in our field
 and rather well-defined PL sequences (see Figure~\ref{fig:wesen}), we can
 calculate the distance to Leo~A using their photometric and pulsational
 properties. As in \citetalias{ber10}, we used both the $P-W_I$ relations for F
 and FO Cepheids, and the period-luminosity-color (PLC) relation for SO Cepheids
 \citep{bon01}. In addition, we also include the distance estimates using the
 recent PL relation of \citet{tam11}, derived specifically for metal-poor
 galaxies from short-period SMC Cepheids.

 The P$-$W$_I$ relations were derived in \citetalias{ber10} by combining the
 LMC and SMC Cepheids from the OGLE-II database \citep{uda99a,uda99b} assuming
 $\Delta(m-M)_0=$~0.51 \citep{uda99c} and an LMC distance modulus of
 (m$-$M)$_{\rm LMC,0}$=18.515$\pm$0.085 \citep{cle03}. We found
\begin{displaymath}
     W_{I,F}  = -3.435(\pm0.007)\ {\rm log}\ P_F - 2.540(\pm0.006),
\end{displaymath}
\begin{displaymath}
     W_{I,FO} = -3.544(\pm0.013)\ {\rm log}\ P_{FO} - 3.067(\pm0.007),
\end{displaymath}
 for the fundamental and first-overtone Cepheids, respectively. We also showed
 that using the F relation from \citet{nge09}, derived using only the LMC
 Cepheids, lead to the same IC\,1613 distance within the uncertainties.

 To take into account the possible change in the PL relations depending on
 metallicity, \citet{tam11} derived new PL and period-color (PC) relations using
 the metal-poor Cepheids of the SMC from the OGLE-II database. The PL
 (and PC) relations of both F and FO pulsators seem to present a break, so they
 fitted two linear regressions, treating the position of the break as a free
 parameter. For the F and FO Cepheids, they find the break at log~$P$ = 0.55 and
 0.40, respectively. A single F Cepheid of our sample has a period longer than
 the break, so we simply rejected it and used the PL relations for periods
 shorter than the breaks.

 Finally, the PLC relation for SO Cepheids was derived by \citet{bon01} from
 their pulsational models. Due to the low amplitude of the SO light curves
 and very narrow instability strip, the uncertainties in distance are relatively
 small.

 We note that the relations of \citet{tam11} were calibrated assuming an SMC
 distance modulus of (m$-$M)$_{\rm SMC,0}$=18.93, while those of \citet{bon01}
 applied to SMC Cepheids yielded (m$-$M)$_{\rm SMC,0}$=19.11. For consistency
 with the distances determined in \citetalias{ber10}, we adapted the zero-points
 of these relations to give (m$-$M)$_{\rm SMC,0}$=19.03.

\begin{figure}
 \includegraphics[width=8cm]{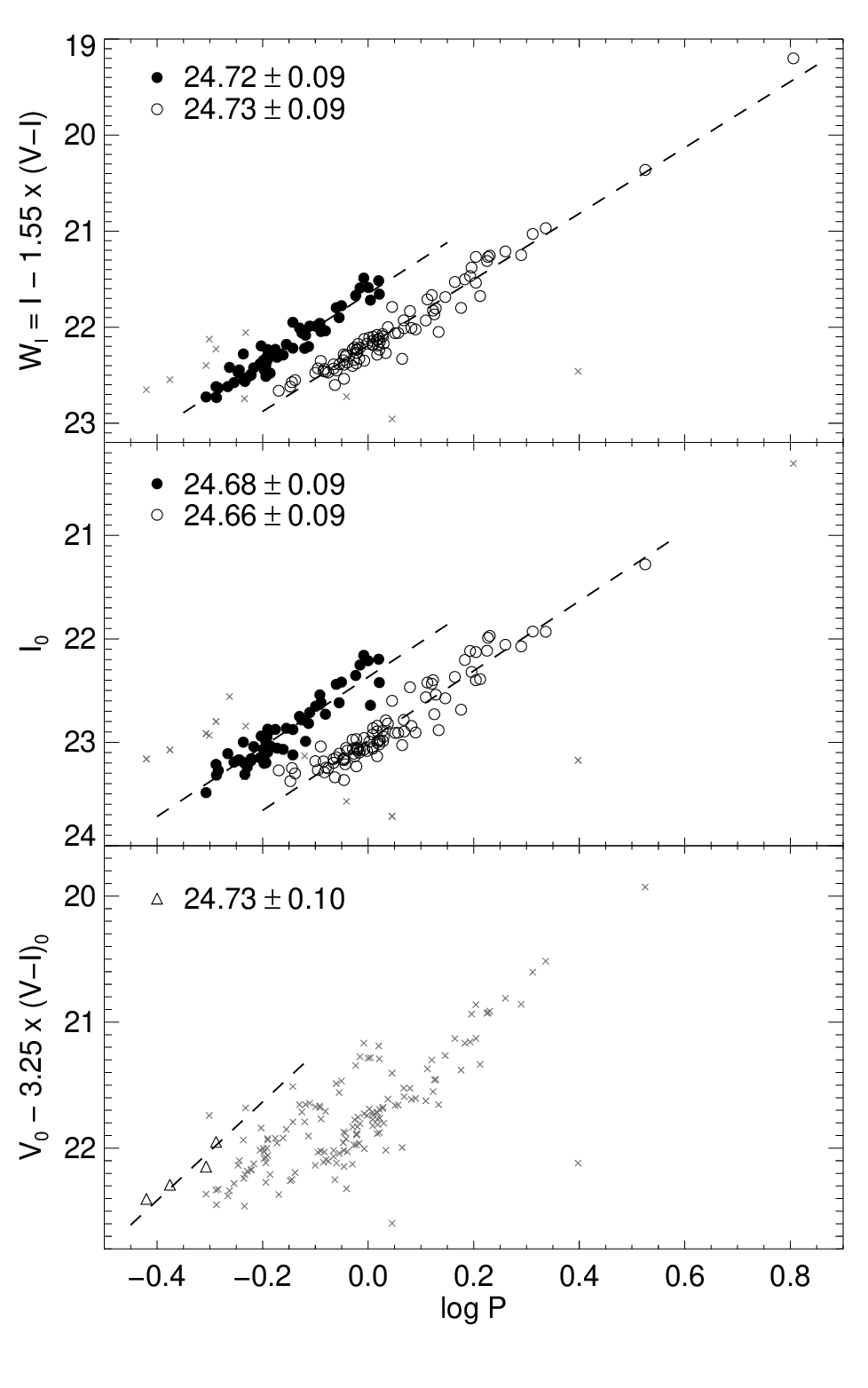}
 \caption{Period-luminosity diagrams for Cepheids in Leo~A. The distance was
 determined using the P$-$W$_I$ relations for F and FO Cepheids of
 \citetalias{ber10} (top), P$-$M$_I$ relations for F and FO Cepheids of
 \citet[middle]{tam11}, and the PLC for SO Cepheids from \citet[bottom]{bon01}.
 In each case, open circles, filled circles, and triangles show the F, FO, and
 SO Cepheids used for the fits, while the gray crosses represent the Cepheids
 that were rejected. The linear fits are shown as dashed lines, and the
 corresponding dereddened distance moduli are listed in the top of each panel.}
 \label{fig:dist}
\end{figure}

 Figure~\ref{fig:dist} summarizes the Cepheid-based distance calculations. The
 panels show the P$-$W$_I$ (top) and P$-$M$_I$ (middle) relations for F and
 FO Cepheids, and the PLC for SO Cepheids (bottom).
 The distance modulus obtained for each fit, corrected for foreground reddening
 (E(B$-$V)=0.021; \citealt{sch98}), is indicated in the top of each panel.
 We find that all the methods give values consistent with each other. A straight
 average of these measurements yield (m$-$M)$_{\rm Leo~A,0}$~=~24.70, with a
 standard deviation of $\sigma=$~0.03.
 This is also in very good agreement with the Cepheid distance of
 \citet{tam11}, after taking into account the correction for the different
 assumed SMC distances.

 Interestingly, while all the Cepheid-based distances agree with each other,
 they are somewhat longer than the distance obtained from the TRGB
 ($\Delta(m-M)_0\sim$~0.13) and from the RR Lyrae stars
 ($\Delta(m-M)_0\sim$~0.18). We do not believe this is an issue with the
 calibration from the {\it HST} flight system to the Johnson-Cousin system: as
 shown in Figure~\ref{fig:wesen}, even in the {\it HST} bands the Cepheids of
 Leo~A are fainter than expected if the galaxy is located 0.1 mag further than
 IC\,1613 as indicated by the difference in TRGB luminosity.
 Likewise, the luminosity of the Cepheid sample is unlikely to be biased toward
 fainter magnitudes due to the presence of a significant population of AC. The
 PL relations shown in Figure~\ref{fig:dist} are very tight, and apparently
 linear over the whole range of periods. Therefore, even if some AC are present
 in the sample, they are indistinguishable from the classical Cepheids and would
 not affect the distance measurements.
 This may indicate that the Cepheids in Leo~A are intrinsically fainter at a
 given period than in the LMC, SMC, and IC\,1613. This is discussed further in
 Section~\ref{sec:discuss2}.

\section{Discussion}\label{sec:discuss}

\subsection{Metallicity Evolution}\label{sec:discuss1}

 We have calculated the distance to Leo~A using several methods based on the
 properties of the Cepheids and RR~Lyrae stars, as well as the luminosity of the
 TRGB. We find a good agreement between the values obtained from the TRGB and
 the RR Lyrae stars ((m$-$M)$_0$=24.57$\pm$0.13 and 24.48$\pm$0.12,
 respectively). These are also consistent with the recent literature value of
 (m$-$M)$_0\sim$~24.5 \citepalias{dol02,sch02} obtained with similar methods.

 The distance inferred from the RR Lyrae stars is quite sensitive to the
 metallicity adopted for these stars, which at the distance of Leo~A cannot be
 directly measured. However, the SFH of Leo~A calculated by Hidalgo et
 al.\ (2013, in preparation) from CMD-fitting provides the age-metallicity
 relation. It shows a very low metallicity at early times (Z$\sim$0.0002,
 [Fe/H]$\sim-$2.0), followed by a steady metal enrichment until the present time
 when the metallicity derived is in fair agreement with the value inferred from
 \hii\ regions.

 The low metallicity for the old population is also required to explain several
 features of the observed CMD, such as the position of the HB and RC, and to
 obtain a good agreement between the observed HB and theoretical models. Close
 examination of the RC in Figure~\ref{fig:tracks} reveals a thin, almost
 vertical group of stars at (F475W$-$F814W)$\sim$1.1 and F814W$\sim$24, slightly
 separated from the blue side of the RC. This suggests the presence of a small
 population of old stars, more metal-poor than the bulk of the RC stars.

\begin{figure}
 \includegraphics[width=8.5cm]{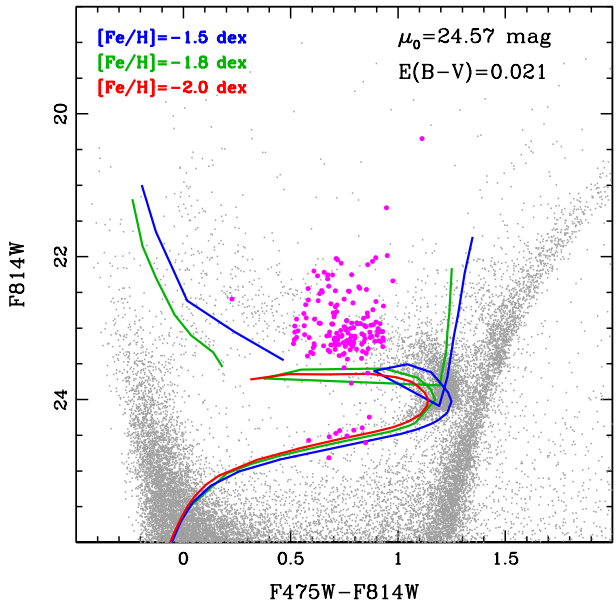}
 \caption{CMD showing the observed variable (magenta filled circles) and
 non-variable (gray dots) stars. Theoretical ZAHeBs are shown for three different
 metallicities. The lines at (F475W$-$F814W)$\sim$0 indicate the
 maximum extent of the blue loops.}
 \label{fig:zahb}
\end{figure}

 To illustrate this, in Figure~\ref{fig:zahb} we show the CMD of Leo~A where
 theoretical ZAHeB of various metallicities have been overlain. The
 lines at (F475W$-$F814W)$\sim$0 indicate the maximum extent of the
 blue loops. According to synthetic HB models in optical photometric bands, the
 ZAHeB locus is expected to be about 0.05 mag fainter than the lower envelope of
 the observed HB distribution \citep[see][]{cas04,cat04,cas13}. This means that
 the best fit is provided by the [Fe/H]$\sim-$2.0 ZAHeB.
 At this metallicity, the ZAHeB also provides a very good
 fit to the ``blue RC''. On the other hand, the young stars of the blue loops
 are better represented by the higher metallicity, in agreement with the
 spectroscopic value of the \hii\ regions. This suggests that the metallicity of
 Leo~A has been increasing from [Fe/H]$\sim-$2 to [Fe/H]$\sim-$1.5 (i.e.\ a
 factor $\sim$3) over the past 10~Gyr or so.

\subsection{Metallicity Dependence of the Cepheid PL Relation}\label{sec:discuss2}

 The reason for the slightly longer distance obtained from the Cepheids
 ($\sim$0.13) is not clear. Given the extremely low metallicity of Leo~A it may
 also be explained by a metallicity effect, although the dependence on
 metallicity of the slopes and zero-points of PL relations is still debated
 \citep[see, e.g.,][]{sak04,rom08,san09,sto11}.
 Figure~\ref{fig:wesen} indicates that the Cepheids of Leo~A are fainter than
 those in IC\,1613, if we assume that the former are located 0.1 mag further
 than the latter as indicated by the difference in TRGB luminosity. Likewise,
 Figure~\ref{fig:pl} suggests that the shortest-period SMC Cepheids are, at a
 given period, slightly fainter than their more metal-rich LMC equivalent (see
 e.g.\ at log P~$\sim-$0.1). The luminosity difference between the short-period
 Cepheids of Leo~A and the LMC is also significant. This may indicate that the
 luminosity of the short-period Cepheids decreases with decreasing metallicity,
 leading to a longer distance modulus than that obtained from the TRGB and the
 RR~Lyrae stars.

 In fact, such a trend with metallicity of the difference between the TRGB and
 Cepheid distances has been suggested before \citep{ken98,sak04}, in the sense
 that the distance modulus obtained from the Cepheids appeared increasingly
 longer than the TRGB distance modulus at decreasing metallicity. However, these
 results have recently been contested by \citet[][see also
 \citealt{bon08}]{riz07} who found no significant trend for the same sample of
 galaxies \citep[see also][]{sal98}.
 While the origin of the discrepancy between the \citet{sak04} and \citet{riz07}
 measurements is not discussed in the latter paper, it suggests that if the
 luminosity of the Cepheids does depend on metallicity the effect is smaller
 than the uncertainties currently affecting the measurements. For the sample of
 galaxies presented by \citet{riz07} -- covering almost two orders of magnitude
 in nebular oxygen abundance -- the difference between TRGB and Cepheid
 distance estimates has a standard deviation of 0.1, similar to the typical
 distance uncertainty to nearby galaxies.

 The discrepancy between TRGB and Cepheid distances could also be unrelated to
 the stellar metallicity. Other possible factors affecting the distances
 obtained from Cepheids include the possible non-linearity of the PL relations
 \citep[e.g.][]{nge09}, the intrinsic color of the Cepheids \citep{san08}, as
 well as the effects of stellar crowding and the properties of the intervening
 dust.
 Therefore, while the short-period Cepheids of Leo~A appear to be intrinsically
 fainter than their equivalent in IC\,1613 and the Magellanic Clouds, our data
 is not sufficient to conclusively claim the detection of a metallicity effect.
 On the other hand, the large number of Cepheids, combined with the fact that
 the metallicity of Leo~A is significantly lower than any galaxy in the
 \citet{riz07} sample, make it a valuable galaxy when studying the metallicity
 dependencies of the variable star PL relationships.

\section{Summary and Conclusions}\label{sec:conclusions}

 We have analyzed time series HST/ACS data of the dIrr galaxy Leo~A with the
 goal of searching for variable stars. Thanks to the significant depth of these
 data (i.e.\ $>$90 percent complete 2~mag below the HB), we have found 156
 Cepheids and 10 RR~Lyrae stars, as well as 33 candidate variables of various
 types. The number of RR~Lyrae stars is very low compared to any other galaxies
 where these variables have been searched for, but is not unexpected given the
 insignificant fraction of old stars ($\ga$8~Gyr) in this galaxy \citep[$\la$10
 percent;][S.\ Hidalgo et al.\ 2013, in preparation]{col07}. The classification
 of Leo~A in one of the Oosterhoff groups from the properties of its RR~Lyrae
 stars is not settled: while the mean periods and number ratio clearly indicate
 a Oosterhoff type~II -- which is uncommon in Local Group galaxies -- the
 location of these stars in the period-amplitude diagram suggests it is
 Oo-Intermediate. Follow-up data covering a larger FOV will be necessary to
 increase the sample.

 The numerous Cepheids generally have very short periods, with the mode of the
 distributions at $\sim$1 and $\sim$0.6 day for F and FO pulsators, respectively.
 From their position on the PL diagram and light-curve
 morphology, we classify 91, 58, and 4 Cepheids as fundamental, first-overtone,
 and second-overtone mode CC, respectively, and two as population II Cepheids.
 We have showed, however, that some of the fainter CC may actually be AC;
 neither the sparse light curves nor their location in the PL diagrams allowed
 us to distinguish them convincingly.

 We have calculated the distance to Leo~A using the TRGB luminosity and various
 methods based on the photometric and pulsational properties of the Cepheids and
 RR~Lyrae stars. Our best distance based on the TRGB method is
 (m$-$M)$_0=$~24.57$\pm$0.13, corresponding to 820$\pm$50~kpc. This distance is
 in good agreement, within the uncertainties, with that obtained from the
 RR~Lyrae stars ((m$-$M)$_0$=24.48$\pm$0.12) and literature values. The distance
 obtained from the Cepheid PL relations, however, is somewhat larger than these
 values, (m$-$M)$_0$=24.70$\pm$0.10.
 This may indicate that the short-period Cepheids of Leo~A are
 intrinsically fainter than their equivalent in IC\,1613 and the Magellanic
 Clouds.
 In any case, the very low metallicity of Leo~A and its substantial populations
 of RGB stars, Cepheids, and RR~Lyrae stars make it a valuable galaxy to
 investigate the possible metallicity dependencies of the variable star PL
 relationships.

 Finally, while the current metallicity of Leo~A is very low -- the lowest
 within the star forming galaxies of the LG -- the morphology of the HB and RC
 in the CMD require that it was even lower about 10~Gyr ago. This suggests that
 the metallicity of Leo~A has increased from [Fe/H]$\sim-$2 to [Fe/H]$\sim-$1.5
 (i.e.\ a factor $\sim$3) over the past 10~Gyr or so.

\section*{acknowledgments}

\small
The authors are grateful to the anonymous referee for thoughtful comments that
helped improve the manuscript.
Support for this work was provided by a rolling grant from the UK Science and
Technology Facilities Council, the IAC (grants 310394 and P/301204), and the
Education and Science Ministry of Spain (grants AYA2007-3E3506 and
AYA2010-16717).
\normalsize

\appendix
\section{Finding Charts}

\begin{figure*}
 \includegraphics[width=16.5cm]{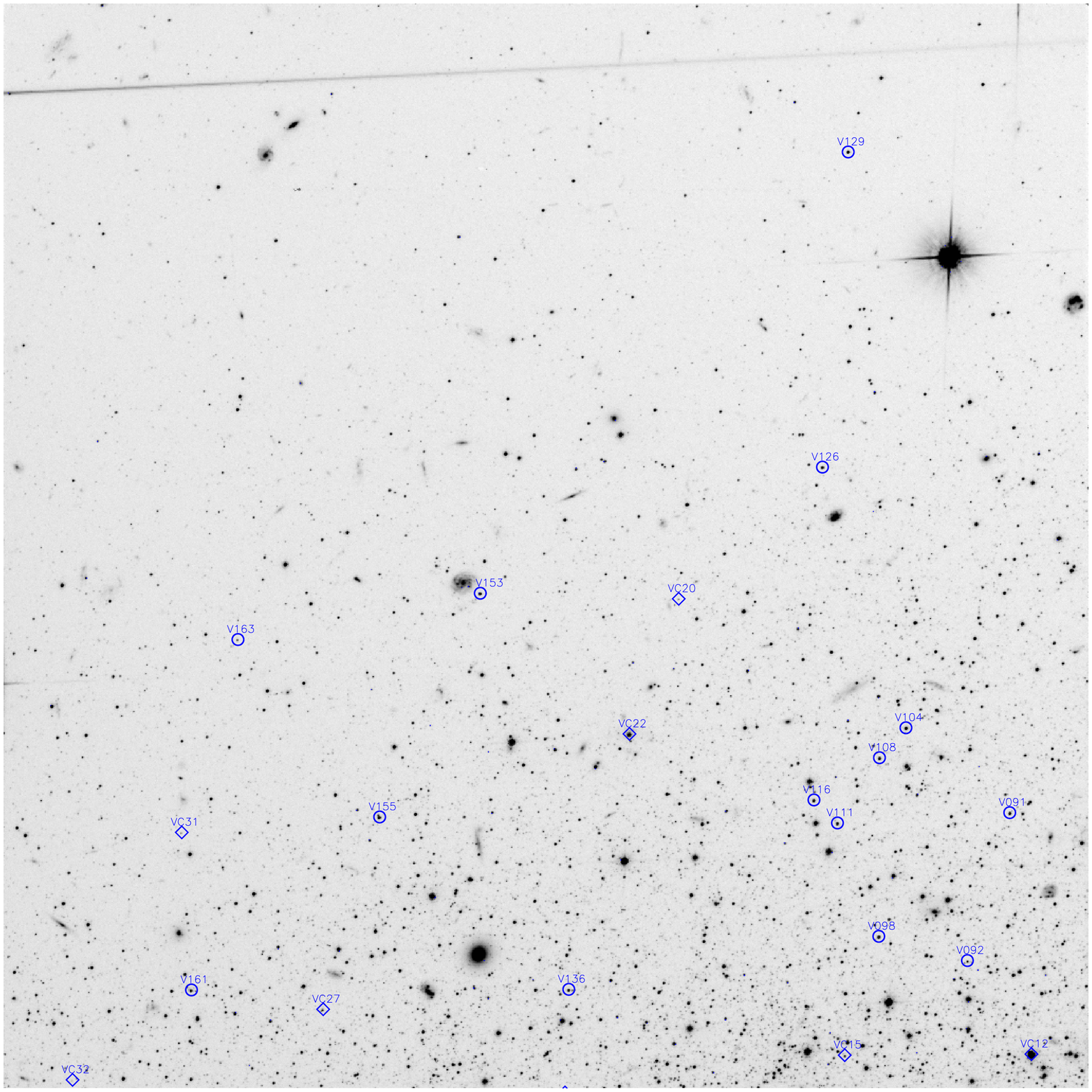}
 \caption{Finding chart for variables (circles) and candidate variables
 (diamonds) in the northeast quadrant. The full set of high-resolution finding
 charts is available as Supporting Information with the online version of the
 paper.}
 \label{fig:fc1}
\end{figure*}

 The finding charts for the whole sample of variable stars are presented
 in the electronic version of this paper (Figure~\ref{fig:fc1}).

\begin{table*}
 \centering
 \scriptsize
 \begin{minipage}{130mm}
  \caption{Cross-Identification with Previously Known Variables in Leo\,A.\label{tab:tab5}}
  \begin{tabular}{@{}rccccccccc@{}}
  \hline
ID & Period & Type\footnote{$ab$: fundamental mode RR~Lyrae star; F, FO: fundamental mode and first-overtone Classical Cepheids; P2C: population II Cepheid; LPV: long period variable; MSV: main-sequence variable.} & 
ID$_{D02}$ & Period & Type\footnote{Type from \citetalias{dol02} -- RR: RR~Lyrae star; C-FM, C-FO, C-SO: fundamental mode, first-, and second-overtone Classical Cepheids; C: Cepheid of unknown type.} & ID$_{G07}$ & Period & ID$_{H94}$ & Period \\
   & (days) &      &            & (days) &      &            & (days) &            & (days) \\
 \hline
V002 & 0.935 &  F   & C2-V24 & 0.47 $\pm$ 0.02 & C-FO/SO &   	  -	    &	-    &  -  &	-     \\
V004 & 0.934 &  F   & C2-V55 & 0.94 $\pm$ 0.03 &    C    &   	  -	    &	-    &  -  &	-     \\
V005 & 1.63  &  F   & C2-V39 & 1.69 $\pm$ 0.05 &   C-FM  &   	 LM53	    &  2.674 &  -  &	-     \\
V007 & 1.60  &  F   & C2-V08 & 1.70 $\pm$ 9.99 &    C    &   	  -	    &	-    &  -  &	-     \\
V008 & 0.542 &  FO  & C2-V27 & 0.55 $\pm$ 0.02 &   C-FO  &   	  -	    &	-    &  -  &	-     \\
V009 & 2.5   & P2C  & C2-V52 & 0.70 $\pm$ 0.06 &    C    &   	  -	    &	-    &  -  &	-     \\
V010 & 0.635 &  FO  & C2-V20 & 0.65 $\pm$ 0.03 &   C-FO  &   	  -	    &	-    &  -  &	-     \\
V012 & 0.570 &  FO  & C2-V03 & 0.60 $\pm$ 0.02 &    C    &   	  -	    &	-    &  -  &	-     \\
V013 & 1.343 &  F   & C2-V23 & 1.52 $\pm$ 0.20 &   C-FM  &   	  -	    &	-    &  -  &	-     \\
V015 & 0.645 &  FO  & C2-V21 & 0.64 $\pm$ 0.02 &    C    &   	  -	    &	-    &  -  &	-     \\
V017 & 0.643 &  FO  & C2-V59 & 0.64 $\pm$ 0.02 &   C-FO  &   	  -	    &	-    &  -  &	-     \\
V021 & 1.82  &  F   &   -    &        -        &    -    &   LM04/LeoA-15   &  1.813 & V13 &  3.41703 \\
V023 & 0.627 &  FO  & C2-V54 & 0.61 $\pm$ 0.01 &   C-FO  &   	  -	    &	-    &  -  &	-     \\
V024 & 0.947 &  FO  & C2-V38 & 0.47 $\pm$ 0.02 &   C-SO  &   	 LL14	    & 18.920 &  -  &	-     \\
V025 & 1.40  &  F   & C2-V37 & 1.28 $\pm$ 0.12 &   C-FM  &   	 LL11	    & 57.912 &  -  &	-     \\
V028 & 1.09  &  F   & C2-V05 & 0.54 $\pm$ 0.02 &   C-FO  &   	  -	    &	-    &  -  &	-     \\
V029 & 0.587 & $ab$ & C2-V66 & 0.54 $\pm$ 0.03 &    RR   &   	  -	    &	-    &  -  &	-     \\
V031 & 1.327 &  F   & C2-V25 & 1.39 $\pm$ 0.15 &   C-FM  &   	 LM10	    &  4.008 & V10 & 13.0040  \\
V035 & 0.981 &  FO  &   -    &        -        &    -    &   	 LS01	    &  0.981 & V9  &  2.6704  \\
V036 & 1.335 &  F   & C2-V09 & 1.22 $\pm$ 0.18 &   C-FO  &   	  -	    &	-    &  -  &	-     \\
V037 & 1.525 &  F   &   -    &        -        &    -    &   LM17/LeoA-22   &  1.524 &  -  &	-     \\
V038 & 0.871 &  F   & C2-V49 & 0.86 $\pm$ 0.02 &   C-FM  &   	  -	    &	-    &  -  &	-     \\
V039 & 0.814 &  FO  & C2-V19 & 0.82 $\pm$ 0.04 &   C-FO  &   	  -	    &	-    &  -  &	-     \\
V040 & 0.607 &  FO  & C2-V26 & 0.61 $\pm$ 0.02 &   C-FO  &   	  -	    &	-    &  -  &	-     \\
V041 & 0.633 &  FO  & C2-V13 & 0.64 $\pm$ 0.01 &   C-FO  &   	  -	    &	-    &  -  &	-     \\
V045 & 0.953 &  F   & C2-V50 & 0.49 $\pm$ 0.01 &   C-FO  &   	  -	    &	-    &  -  &	-     \\
V053 & 1.08  &  F   &   -    &        -        &    -    &   	 LS39	    &  0.188 &  -  &	-     \\
V055 & 0.70  &  FO  & C2-V34 & 0.68 $\pm$ 0.01 &   C-FO  &   	  -	    &	-    &  -  &	-     \\
V059 & 0.515 &  FO  &   -    &        -        &    -    &   	 LL02	    & 65.446 &  -  &	-     \\
V060 & 0.965 &  FO  &   -    &        -        &    -    &   LS26/LeoA-21   &  0.965 &  -  &	-     \\
V061 & 2.170 &  F   &   -    &        -        &    -    & 	 LM16	    &  2.164 &  -  &	-     \\
V062 & 0.640 &  FO  & C2-V61 & 0.64 $\pm$ 0.01 &   C-FO  & 	  -	    &	-    &  -  &	-     \\
V064 & 1.32  &  F   &   -    &        -        &    -    & 	 LM15	    &  1.320 &  -  &	-     \\
V067 & 3.35  &  F   &   -    &        -        &    -    & LM03/V03/LeoA-11 &  3.354 &  -  &	-     \\
V073 & 1.95  &  F   &   -    &        -        &    -    & 	 LM13	    &  1.943 &  -  &	-     \\
V082 & 1.57  &  F   & C2-V43 & 1.61 $\pm$ 0.05 &   C-FM  & 	 LM39	    &  1.562 &  -  &	-     \\
V085 & 0.624 &  FO  & C2-V53 & 0.63 $\pm$ 0.03 &    C    & 	  -	    &	-    &  -  &	-     \\
V088 & 1.56  &  F   &   -    &        -        &    -    & LM07/V07/LeoA-9  &  1.564 & V7  &  0.54837 \\
V089 & 0.890 &  FO  &   -    &        -        &    -    & 	 LM54	    &  8.390 &  -  &	-     \\
V090 & 1.17  &  F   & C2-V69 & 1.15 $\pm$ 0.04 &   C-FM  & 	  -	    &	-    &  -  &	-     \\
V093 & 1.60  &  F   & C2-V42 & 1.73 $\pm$ 0.23 &   C-FO  & LM02/V06/LeoA-20 &  1.607 &  -  &	-     \\
V098 & 1.022 &  F   & C2-V11 & 0.51 $\pm$ 0.01 &   C-SO  & 	 LM47	    &  1.021 &  -  &	-     \\
V100 & 0.516 &  FO  & C2-V36 & 0.52 $\pm$ 0.02 &    C    & 	  -	    &	-    &  -  &	-     \\
V104 & 1.065 &  F   & C2-V04 & 0.53 $\pm$ 0.02 &    C    & 	  -	    &	-    &  -  &	-     \\
V107 & 1.69  &  F   & C2-V22 & 1.67 $\pm$ 0.13 &   C-FM  & 	 LM11	    &  1.690 &  -  &	-     \\
V111 & 0.90  &  F   & C2-V07 & 0.87 $\pm$ 0.02 &    C    & 	  -	    &	-    &  -  &	-     \\
V114 & 1.46  &  F   & C2-V31 & 1.46 $\pm$ 0.20 &   C-FM  & 	  -	    &	-    &  -  &	-     \\
V116 & 0.690 &  FO  & C2-V06 & 0.68 $\pm$ 0.01 &   C-FO  & 	  -	    &	-    &  -  &	-     \\
V119 & 1.286 &  F   & C2-V60 & 1.32 $\pm$ 0.11 &   C-FM  & 	 LM14	    &  1.280 &  -  &	-     \\
V122 & 0.810 &  FO  & C2-V32 & 0.78 $\pm$ 0.04 &   C-FO  & 	  -	    &	-    &  -  &	-     \\
V124 & 1.687 &  F   &   -    &        -        &    -    &     LM09/V02     &  1.685 &  -  &	-     \\
V126 & 0.983 &  F   &   -    &        -        &    -    & 	 LS35	    &  0.450 &  -  &	-     \\
V130 & 1.20  &  F   &   -    &        -        &    -    & 	 LM21	    &  5.848 &  -  &	-     \\
V135 & 6.394 &  F   &   -    &        -        &    -    & LM01/V01/LeoA-5  &  6.490 &  -  &	-     \\
V141 & 0.749 &  FO  & C2-V30 & 0.75 $\pm$ 0.06 &   C-FO  & 	  -	    &	-    &  -  &	-     \\
V145 & 2.05  &  F   & C2-V64 & 2.01 $\pm$ 9.99 &    C    &     LM06/V04     &  2.048 &  -  &	-     \\
V147 & 0.813 &  F   & C2-V46 & 0.60 $\pm$ 0.02 &   C-FO  & 	  -	    &	-    &  -  &	-     \\
V148 & 1.50  &  F   &   -    &        -        &    -    & 	 LM45	    &  1.499 &  -  &	-     \\
V153 & 0.652 &  FO  & C2-V02 & 0.65 $\pm$ 0.02 &   C-FO  & 	  -	    &	-    &  -  &	-     \\
V155 & 0.775 &  FO  & C2-V10 & 0.79 $\pm$ 0.04 &   C-FO  & 	 LS53	    &  0.131 &  -  &	-     \\
V157 & 1.68  &  F   & C2-V58 & 1.46 $\pm$ 0.17 &   C-FM  &     LM05/V05     &  1.685 &  -  &	-     \\
V158 & 1.052 &  F   &   -    &        -        &    -    & 	 LL05	    & 10.391 &  -  &	-     \\
V159 & 0.580 &  FO  & C2-V48 & 0.59 $\pm$ 0.03 &   C-FO  & 	  -	    &	-    &  -  &	-     \\
V160 & 0.870 &  FO  &   -    &        -        &    -    & 	 LS29	    &  0.465 &  -  &	-     \\
V162 & 0.590 &  FO  & C2-V35 & 0.39 $\pm$ 0.01 & C-FO/SO &        -         &	-    &  -  &	-     \\
V164 & 0.667 &  FO  & C2-V45 & 0.66 $\pm$ 0.02 &   C-FO  & 	  -	    &	-    &  -  &	-     \\
V165 & 0.795 &  FO  & C2-V28 & 0.80 $\pm$ 0.04 &   C-FO  & 	 LM43	    &  3.897 &  -  &	-     \\
V166 & 0.586 &  FO  & C2-V29 & 0.59 $\pm$ 0.02 &   C-FO  & 	  -	    &	-    &  -  &	-     \\
VC05 &   -   & LPV  &   -    &        -        &    -    &     LeoA-14      & 170.70 & V11 &  2.0064  \\
VC09 &   -   & LPV  &   -    &        -        &    -    &     LeoA-12      & 268.42 &  -  &	-     \\
VC12 &   -   & LPV  &   -    &        -        &    -    &     LeoA-10      & 919.28 &  -  &	-     \\
VC30 &   -   & MSV  &   -    &        -        &    -    & 	 LS56	    &  0.269 &  -  &	-     \\
\hline
\end{tabular}
\end{minipage}
\end{table*}

\label{lastpage}

\end{document}